\documentclass[aps,prd,nofootinbib,superscriptaddress,prd,aps,floatfix]{revtex4}
\usepackage[latin9]{inputenc}
\usepackage{textcomp}
\usepackage{amsmath}
\usepackage{amssymb}
\usepackage{graphicx}
\usepackage[unicode=true,
 bookmarks=false,
 breaklinks=false,pdfborder={0 0 1},backref=section,colorlinks=false]
 {hyperref}
\hypersetup{
 colorlinks,bookmarksopen,bookmarksnumbered,linkcolor=blu,pdfstartview=FitH,urlcolor=rossos,citecolor=rossos}

\usepackage{textcomp}\usepackage{color}\usepackage{textcomp}\usepackage{amsfonts}\usepackage{graphics}\usepackage{epstopdf}\usepackage{slashed}\usepackage{multirow}\usepackage{adjustbox}\usepackage{lipsum}\usepackage{rotating}\usepackage{xcolor}\usepackage{wrapfig}\usepackage{epsfig}
\usepackage[justification=raggedright, margin=5mm,font={sf,small},labelfont=bf, format=plain]{caption}

\definecolor{rosso}{cmyk}{0,1,1,0.4}
\definecolor{rossos}{cmyk}{0,1,1,0.55}
\definecolor{rossoc}{cmyk}{0,1,1,0.2}
\definecolor{blu}{cmyk}{1,1,0,0.3}
\definecolor{blus}{cmyk}{1,1,0,0.6}
\definecolor{bluc}{cmyk}{1,1,0,0.1}
\definecolor{verde}{cmyk}{0.92,0,0.59,0.25}
\definecolor{verdec}{cmyk}{0.92,0,0.59,0.15}
\definecolor{verdes}{cmyk}{0.92,0,0.59,0.4}

\begin{document}

\title{\color{bluc}Radiative Neutrino Mass \& Majorana Dark Matter within
an Inert Higgs Doublet Model}

\author{Amine Ahriche}
\email{aahriche@daad-alumni.de}

\affiliation{Department of Physics, University of Jijel, PB 98 Ouled Aissa, DZ-18000
Jijel, Algeria.}

\affiliation{The Abdus Salam International Centre for Theoretical Physics, Strada
Costiera 11, I-34014, Trieste, Italy.}

\author{Adil Jueid}
\email{adil.jueid@sjtu.edu.cn}

\affiliation{INPAC, Shanghai Key Laboratory for Particle Physics and Cosmology,
Department of Physics and Astronomy, Shanghai Jiao Tong University,
Shanghai 200240, China.}

\affiliation{Département des Mathématiques, Faculté des Sciences et Techniques,
Université Abdelmalek Essaadi, Tanger, Maroc.}

\author{Salah Nasri}
\email{snasri@uaeu.ac.ae}

\affiliation{Department of physics, United Arab Emirates University, Al-Ain, UAE.}

\affiliation{The Abdus Salam International Centre for Theoretical Physics, Strada
Costiera 11, I-34014, Trieste, Italy.}
\begin{abstract}
We consider an extension of the standard model (SM) with an inert
Higgs doublet and three Majorana singlet fermions to address both
origin and the smallness of neutrino masses and dark matter (DM) problems.
In this setup, the lightest Majorana singlet fermion plays the role
of DM candidate and the model parameter space can be accommodated
to avoid different experimental constraints such as lepton flavor
violating processes and electroweak precision tests. The neutrino
mass is generated at one-loop level a la Scotogenic model and its
smallness is ensured by the degeneracy between the CP-odd and CP-even
scalar members of the inert doublet. Interesting signatures at both
leptonic and hadronic colliders are discussed. 
\end{abstract}

\pacs{04.50.Cd, 98.80.Cq, 11.30.Fs.}
\maketitle

\section{Introduction}

The Discovery of the Higgs particle at the LHC in 2012~\cite{Aad:2012tfa},
validated the standard model (SM) of particle physics. Although the
SM has been very successful in describing and explaining the non-gravitational
fundamental interactions between elementary particles, there are still
open questions that it does not answer. For instance, the observation
of neutrino oscillations in solar, atmospheric, reactor and accelerator
experiments confirmed that neutrino have tiny non zero mass, unlike
in the SM where they are strictly massless. Another issue that the
SM does not account for is the existence of dark matter (DM) inferred
from different astronomical and cosmological observations. Thus, going
beyond the SM seems to be necessary in order to address these problems.

The most popular mechanism for explaining the smallness of neutrino
mass is the see-saw mechanism, where a massive right handed neutrino
(RHN) couples to the lepton and the Higgs doublets fields, and which
at low energies induces the effective dimension five operator 
\begin{eqnarray}
{\cal L}_{\text{eff}}^{(5)}=\frac{\kappa_{ij}}{\Lambda}~(\bar{L_{i}^{c}}~i\sigma_{2}~H)(L_{j}^{T}~i\sigma_{2}~H)+h.c,
\end{eqnarray}
where $\Lambda$ is the scale of new physics which is of order the
RH neutrino mass, $\kappa_{ij}$ is a product of Yukawa couplings,
$L$ and $L^{c}$ are the Lepton $SU(2)_{L}$ doublet and its charge
conjugate, respectively, and $H$ is the Higgs doublet. However, on
the basis of naturalness, the scale $\Lambda$ needs to be of order
the GUT scale, making it impossible to probe it in high energy laboratory
experiments. One way to lower the scale of the new physics is by generating
neutrino masses radiatively where their smallness can be naturally
explained by the loop suppression factor(s) and the Yukawa couplings.
This can be realized at one loop~\cite{Zee:1985rj}, two loops~\cite{Zee-86},
three loops~\cite{Aoki:2008av,knt}, or four loops~\cite{Nomura:2016fzs}.
In~\cite{knt3} it has been shown that there is a class of three
loop neutrino mass generation models, where the new particles that
enter in the neutrino loop can be promoted to triplets~\cite{knt3},
quintuplet~\cite{knt5}, septuplets~\cite{knt7}, and within a scale-invariant
framework~\cite{Ahriche:2015loa}. An interesting features of these
class of models is that the scale of new physics can be of order TeV,
making them testable at high energy colliders~\cite{hna,Cheung:2004xm}
(For a review, see~\cite{Cai:2017jrq}). Moreover, the electroweak
phase transition can be strongly first order, an essential ingredient
for successful electroweak baryogenesis~\cite{EWPhT}.

One of the simplest extensions of the scalar sector of the SM is the
inert Higgs doublet model (IHDM) where one add an extra scalar doublet,
$\Phi$, and assumes that there is a discrete $Z_{2}$ symmetry under
which $\Phi$ is odd whereas the SM fields are even ~\cite{Deshpande:1977rw}.
The Phenomenology of the (IHDM) has been studied extensively in the
literature~\cite{Gustafsson:2007pc}. The discrete symmetry $Z_{2}$
has the following features: i) absence of the flavor changing neutral
currents at tree level~\cite{Glashow:1976nt}, ii) the scalar of
the inert doublet does not have interactions with active fermions
and therefore its neutral component, denoted by $H^{0}$, can be a
good candidate for DM. It has been shown that if $H^{0}$ is lighter
than about $\leq50\textrm{ GeV}$, the null results from DM direct
detection experiment implies that annihilation cross section of the
neutral inert scalar into SM light fermions is highly suppressed,
giving too large relic density to what is observed by Planck experiment
\footnote{This can be avoided if there are other annihilation channels as in~\cite{Ahriche:2016cio}.}.
In the mass range $140\textrm{ GeV}<m_{H^{0}}\leq550\textrm{GeV}$,
the annihilation into $W^{+}W^{-}$ is too large, rendering the relic
density too small to be compatible with the astrophysical observations.
There are three mass regions where the inert scalar can be a viable
DM candidate: i) $m_{H^{0}}\simeq m_{h}/2$, with $m_{h}$ is the
SM Higgs mass, corresponding to the annihilation via the s-channel
resonance due to Higgs exchange~\cite{Borah:2017dqx}, ii) $m_{H^{0}}$
around the mass of the $W$ gauge boson, where the annihilation is
into the three-body final state $WW^{*}\rightarrow Wff'$ ~\cite{Honorez:2010re},
and iii) $m_{H^{0}}\sim~\text{TeV}$ with scalar couplings of order
unity~\cite{Choubey:2017hsq}. One possible way to produce the correct
relic density and not being in conflict with DM direct detection experiment
is by extending this model with heavy $SU(2)_{L}$ singlet vector-like
charged leptons~\cite{Borah:2017dqx}.

In the present work, we consider the SM extended by an inert doublet
and three RHNs~\cite{Ma:2006km} in order to address both neutrino
mass and DM problem. We will investigate the case where the lightest
Majorana singlet fermion plays the role of DM candidate instead of
the neutral CP-odd or CP-even scalars\footnote{In this radiative neutrino mass model, the allowed mass range for
inert scalar dark matter remains very similar to the case of pure
IHDM~\cite{Borah:2017dqx}.}. We will show that the model parameter space can be accommodated
to avoid different experimental constraints such as lepton flavor
violating processes and electroweak precision tests. In this setup,
the neutrino mass smallness is ensured by making the splitting between
the CP-even and CP-odd scalars very tiny. We will study different
phenomenological aspects of the model and investigate the different
regions of the parameter space that fulfill various theoretical and
experimental constraints. In addition, we discuss the signatures for
probing the model at both leptonic and hadronic colliders are discussed.

The paper is organized as follows; in section~\ref{sec:model}, we
highlight the model, its parameters and the different constraints
under which the model is subject to. In section~\ref{sec:analysis},
we present some selected results. In section \ref{sec:collider},
we discuss briefly the possible signatures of this model at hadronic
and leptonic collider machines. In section~\ref{sec:conclusion},
we give our conclusion.

\section{The Model: Parameters and Constraints}

\label{sec:model}

\subsection{The Model and Mass Spectrum}

Here we consider the extension of the SM with an inert doublet $\Phi$
and three singlet Majorana fermions $N_{i}\sim(1,1,0)$, $i=1,2,3$,
both odd under a discrete $Z_{2}$ symmetry. The relevant terms for
generating neutrino mass at one loop are 
\begin{eqnarray}
\mathcal{L} & \supset & h_{ij}\bar{L}_{i}\epsilon\Phi N_{j}+\frac{1}{2}M_{i}\bar{N}_{i}^{C}N_{i}+h.c.,\label{LL}
\end{eqnarray}
where $\bar{L}_{i}$ is the left-handed lepton doublet and $\epsilon=i\sigma_{2}$
is an antisymmetric tensor. The scalar potential can be written as
\begin{eqnarray}
V & = & -\mu_{1}^{2}|H|^{2}+\mu_{2}^{2}|\Phi|^{2}+\frac{\lambda_{1}}{6}|H|^{4}+\frac{\lambda_{2}}{6}|\Phi|^{4}+\lambda_{3}|H|^{2}|\Phi|^{2}+\frac{\lambda_{4}}{2}|H^{\dagger}\Phi|^{2}\nonumber \\
 & + & \frac{\lambda_{5}}{4}\left[(H^{\dagger}\Phi)^{2}+h.c.\right],\label{V}
\end{eqnarray}
and the SM Higgs and the inert doublets can be represented as 
\begin{eqnarray}
H=\left(\begin{array}{c}
G^{+}\\
\frac{1}{\sqrt{2}}(\upsilon+h+iG^{0})
\end{array}\right),~\Phi=\left(\begin{array}{c}
H^{+}\\
\frac{1}{\sqrt{2}}(H^{0}+iA^{0})
\end{array}\right).
\end{eqnarray}

Keeping in mind that all field dependent masses are written in the
from $m_{i}^{2}(h)=\mu_{i}^{2}+\frac{1}{2}\alpha_{i}h^{2}$, the parameters
$\lambda_{1}$ and $\mu_{1}^{2}$ in (\ref{V}) can be eliminated
in favor of the SM Higgs mass and its Vacuum Expectation Value ($\upsilon=246~GeV$),
which is considered at one-loop level as 
\begin{eqnarray}
\lambda_{1} & = & \frac{3m_{h}^{2}}{\upsilon^{2}}-\frac{3}{32\pi^{2}}\sum_{i=\textrm{all}}n_{i}\alpha_{i}^{2}\log\frac{m_{i}^{2}}{m_{h}^{2}},\nonumber \\
\mu_{1}^{2} & = & \frac{1}{6}\lambda_{1}\upsilon^{2}+\frac{1}{32\pi^{2}}\sum_{i=\textrm{all}}n_{i}\alpha_{i}m_{i}^{2}\left(\log\frac{m_{i}^{2}}{m_{h}^{2}}-1\right).\label{1l}
\end{eqnarray}
Here, the one-loop contribution to the effective potential is defined
a la $\overline{DR}$ scheme~\cite{Martin:2001vx}, where the renormalization
scale is taken to be the Higgs mass\footnote{The one-loop corrections in (\ref{1l}) could be important if the
coupling combinations $\alpha_{i}=\lambda_{3},~\lambda_{3}+\lambda_{4}/2\pm\lambda_{5}/2$
are large, and/or the inert scalar doublet scalars are heavy.}. After the spontaneous symmetry breaking, we are left with two CP-even
scalars $(h,H^{0})$, one CP-odd scalar $A^{0}$ and a pair of charged
scalars $H^{\pm}$. Their tree-level masses are given by: 
\begin{eqnarray}
m_{H^{\pm}}^{2} & = & \mu_{2}^{2}+\frac{1}{2}\lambda_{3}\upsilon^{2},~m_{H^{0},A^{0}}^{2}=m_{H^{\pm}}^{2}+\frac{1}{4}\left(\lambda_{4}\pm\lambda_{5}\right)\upsilon^{2}.
\end{eqnarray}

\begin{figure}[!h]
\begin{centering}
\includegraphics[width=0.5\textwidth]{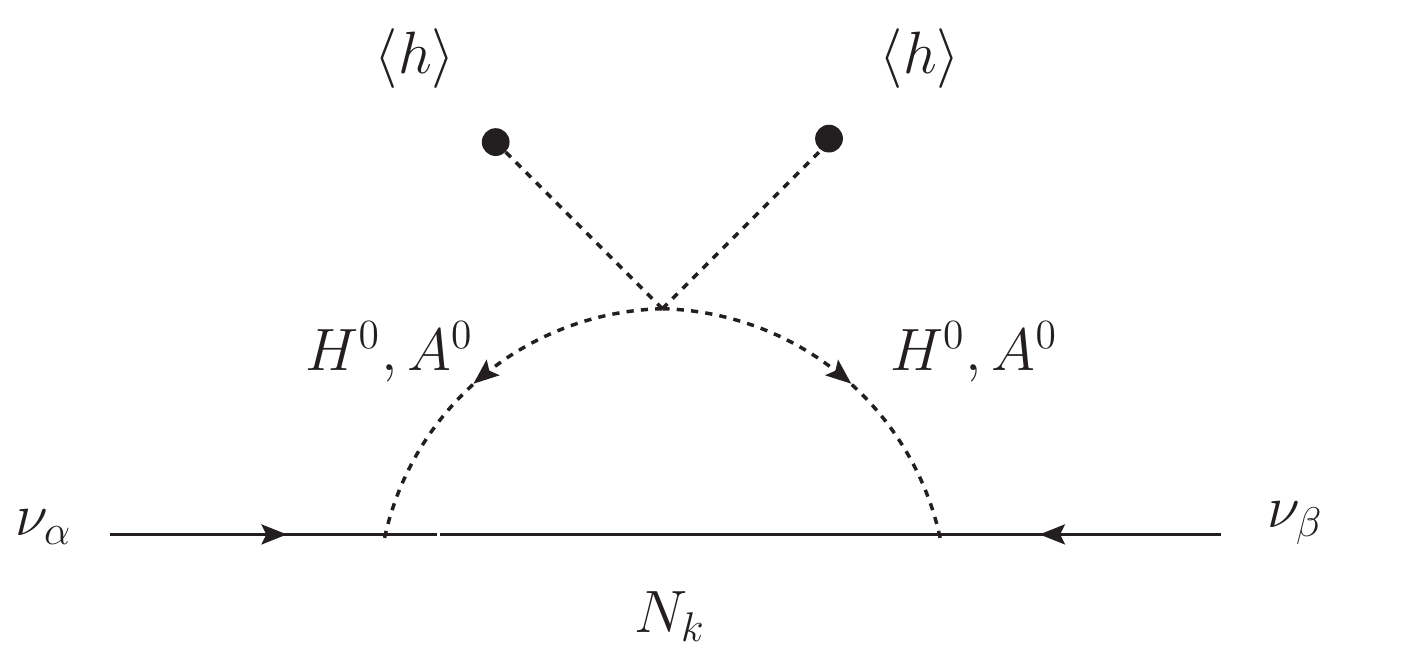} 
\par\end{centering}
\caption{Feynman diagram responsible for the neutrino mass.}
\label{fig:NuM} 
\end{figure}

Note that the $Z_{2}$ symmetry forbids the term $Y_{ij}\bar{L}_{i}HN_{j}$
in the Lagrangian, and hence rendering neutrinos massless at the tree
level. However they get mass at the one loop level (see Fig.~\ref{fig:NuM}),
given\footnote{Note that we use the corrected over-all factor 1/32 ~\cite{Merle:2015gea}
instead of the 1/16 given in~\cite{Ma:2006km}.} 
\begin{eqnarray}
m_{\alpha\beta}^{(\nu)}=\sum_{k}\frac{h_{\alpha k}h_{\beta k}M_{k}}{32\pi^{2}}\left[\frac{m_{H^{0}}^{2}}{m_{H^{0}}^{2}-M_{k}^{2}}\ln\frac{m_{H^{0}}^{2}}{M_{k}^{2}}-\frac{m_{A^{0}}^{2}}{m_{A^{0}}^{2}-M_{k}^{2}}\ln\frac{m_{A^{0}}^{2}}{M_{k}^{2}}\right].\label{Mnu}
\end{eqnarray}
If the mass splitting $m_{H^{0}}^{2}-m_{A^{0}}^{2}=\frac{1}{2}|\lambda_{5}|\upsilon^{2}$
is small with respect to the average $\bar{m}^{2}=(m_{H^{0}}^{2}+m_{A^{0}}^{2})/2$,
which is favored by the Electroweak precision measurements, then the
expression of the mass matrix elements is simplified to 
\begin{eqnarray}
m_{\alpha\beta}^{(\nu)}\simeq\frac{|\lambda_{5}|\upsilon^{2}}{16\pi^{2}}\sum_{k}\frac{h_{\alpha k}h_{\beta k}M_{k}}{\bar{m}^{2}-M_{k}^{2}}\left[1-\frac{M_{k}^{2}}{\bar{m}^{2}-M_{k}^{2}}\ln\frac{\bar{m}^{2}}{M_{k}^{2}}\right].\label{eq:mnu:1}
\end{eqnarray}

The neutrino mass matrix elements in \eqref{eq:mnu:1} can be related
to the elements of the Pontecorvo-Maki-Nakawaga-Sakata (PMNS) mixing
matrix~\cite{Pontecorvo:1967fh} elements. We parametrize the latter
as 
\begin{equation}
U_{\nu}=\left(\begin{array}{ccc}
c_{12}c_{13} & c_{13}s_{12} & s_{13}e^{-i\delta_{d}}\\
-c_{23}s_{12}-c_{12}s_{13}s_{23}e^{i\delta_{d}} & c_{12}c_{23}-s_{12}s_{13}s_{23}e^{i\delta_{d}} & c_{13}s_{23}\\
s_{12}s_{23}-c_{12}c_{23}s_{13}e^{i\delta_{d}} & -c_{12}s_{23}-c_{23}s_{12}s_{13}e^{i\delta_{d}} & c_{13}c_{23}
\end{array}\right)\times U_{m},
\end{equation}
with $\delta_{d}$ the Dirac phase and $U_{m}=\mathrm{diag}(1,\,e^{i\theta_{\alpha}/2},\,e^{i\theta_{\beta}/2})$
encoding the Majorana phase dependence. The shorthand $s_{ij}\equiv\sin\theta_{ij}$
and $c_{ij}\equiv\cos\theta_{ij}$ refers to the mixing angles. For
our numerical scans (discussed below) we fit to the best-fit experimental
values for the mixing angles and mass-squared differences: $s_{12}^{2}=0.320_{-0.017}^{+0.016}$,
$s_{23}^{2}=0.43_{-0.03}^{+0.03}$, $s_{13}^{2}=0.025_{-0.003}^{+0.003}$,
$|\Delta m_{13}^{2}|=2.55_{-0.09}^{+0.06}\times10^{-3}\mathrm{eV}^{2}$
and $\Delta m_{21}^{2}=7.62_{-0.19}^{+0.19}\times10^{-5}\mathrm{eV}^{2}$~\cite{Tortola:2012te}.
Furthermore, we require that the contribution to neutrino-less double
beta decay in this model satisfies the current bound. Within these
ranges, one determines the parameter space where viable neutrino masses
and mixing occur in the model.

\subsection{Theoretical and Experimental Constraints}

Here, we discuss different theoretical and experimental constraints
on the model parameters. \\

\paragraph{Theoretical Constraints:}

The parameters of the scalar potential have to satisfy these theoretical
constraints: 
\begin{itemize}
\item \textbf{Perturbativity}: all the quartic couplings of the physical
fields should be less than $4\pi$, i.e., 
\begin{eqnarray}
\lambda_{1},\lambda_{2},\left\vert \lambda_{3}+\frac{1}{2}\lambda_{4}\right\vert ,\left\vert \lambda_{3}+\frac{1}{2}(\lambda_{4}\mp\lambda_{5})\right\vert \leq4\pi.\label{lm}
\end{eqnarray}
\item \textbf{Vacuum Stability}: the scalar potential is required to be
bounded from below in all the directions of the field space. In both
field planes $h$--$H^{0}$ and $G^{0}$--$A^{0}$ the following
the condition must be satisfied~\cite{Klimenko} 
\begin{equation}
\lambda_{1},\lambda_{2},\lambda_{3}+\frac{1}{2}(\lambda_{4}+\lambda_{5})+\sqrt{\lambda_{1}\lambda_{2}}>0,
\end{equation}
whereas in the plane $H^{^{\pm}}$--$G^{^{\mp}}$, we find 
\begin{equation}
\lambda_{3}+\frac{1}{2}\lambda_{4}+\frac{2}{3}\sqrt{\lambda_{1}\lambda_{2}}>0.
\end{equation}

In addition, we should consider the condition 
\begin{equation}
\frac{\mu_{1}^{2}}{\sqrt{\lambda_{1}}}\geq-\frac{\mu_{2}^{2}}{\sqrt{\lambda_{2}}},
\end{equation}
which is required to guarantee that the inert vacuum is the global
minimum~\cite{Ginzburg:2010wa}. 
\item \textbf{Perturbative unitarity}: We demand that the perturbative unitarity
is preserved in variety of processes involving scalars or gauge bosons
at high energy. At high energies, the equivalence theorem replaces
the $W$ and $Z$ bosons by the Goldstone bosons. Computing the decay
amplitudes for these processes, one finds a set of $4$ matrices with
quartic couplings as their entries~\cite{Akeroyd:2000wc}. The diagonalization
of the scattering matrix gives the following eigenvalues 
\begin{eqnarray}
e_{1,2} & = & \lambda_{3}\pm\frac{\lambda_{4}}{2},~e_{3,4}=\lambda_{3}\pm\frac{\lambda_{5}}{2},~e_{5,6}=\lambda_{3}+\lambda_{4}\pm\frac{3}{2}\lambda_{5},~e_{7,8}=-\frac{\lambda_{1}+\lambda_{2}}{6}\pm\sqrt{\left(\frac{\lambda_{1}+\lambda_{2}}{6}\right)^{2}+\frac{\lambda_{4}^{2}}{4}},\nonumber \\
e_{9,10} & = & -\frac{\lambda_{1}+\lambda_{2}}{2}\pm\sqrt{9\left(\frac{\lambda_{1}-\lambda_{2}}{6}\right)^{2}+\left(2\lambda_{3}+\frac{\lambda_{4}}{2}\right)^{2}},~e_{11,12}=-\frac{\lambda_{1}+\lambda_{2}}{6}\pm\sqrt{\left(\frac{\lambda_{1}-\lambda_{2}}{6}\right)^{2}+\frac{\lambda_{5}^{2}}{4}}.
\end{eqnarray}
We require that the largest eigenvalue of these matrices to be smaller
than $4\pi$. 
\end{itemize}

\paragraph{Experimental Constraints}
\begin{itemize}
\item \textbf{Gauge bosons decay widths}: in order to keep the $W$ and
$Z$ gauge bosons decay modes unmodified, one needs to impose the
following conditions: 
\begin{eqnarray}
m_{H^{0}}+m_{A^{0}}>M_{Z}~,~m_{H^{\pm}}+m_{A^{0}}>M_{W},\nonumber \\
2m_{H^{\pm}}>M_{Z}~,~m_{H^{\pm}}+m_{H^{0}}>M_{W}.
\end{eqnarray}
\item \textbf{Lepton flavor violation (LFV) processes}: in this model, LFV
decay processes arise at one-loop level with the exchange of $H^{\pm}$
and $N_{k}$ particles. The branching ratio of the decay $\ell_{\alpha}\rightarrow\ell_{\beta}+\gamma$
due to the contribution of the interactions (\ref{LL}) is~\cite{Toma:2013zsa}.
\begin{equation}
Br(\ell_{\alpha}\rightarrow\ell_{\beta}+\gamma)=\frac{3\alpha\upsilon^{4}}{32\pi m_{H^{\pm}}^{4}}\left|\sum_{i=1}^{3}h_{\beta i}^{\ast}h_{\alpha i}F\left(M_{i}^{2}/m_{H^{\pm}}^{2}\right)\right|^{2},\label{meg}
\end{equation}
where $\alpha=e^{2}/4\pi$ is the electromagnetic fine structure constant
and $F\left(x\right)=\left(1-6x+3x^{2}+2x^{3}-6x^{2}\log x\right)/6\left(1-x\right)^{4}$.
We will consider also the LFV decays $\ell_{\alpha}\rightarrow\ell_{\beta}\ell_{\beta}\ell_{\beta}$,
where their branching ratio formulas are given in~\cite{Toma:2013zsa}.
In our numerical scan, we will impose all the experimental limits
on both $Br(\ell_{\alpha}\rightarrow\ell_{\beta}+\gamma)$ and $Br(\ell_{\alpha}\rightarrow\ell_{\beta}\ell_{\beta}\ell_{\beta})$
~\cite{Patrignani:2016xqp}. 
\item \textbf{The electroweak precision tests}: while taking $\Delta U=0$
in our analysis, the oblique parameters can be written in our model
as~\cite{Grimus:2008nb} 
\begin{eqnarray}
\varDelta T & = & \frac{1}{16\pi s_{\mathrm{w}}^{2}M_{W}^{2}}\left\{ F\left(m_{H^{0}}^{2},m_{H^{\pm}}^{2}\right)+F\left(m_{A^{0}}^{2},m_{H^{\pm}}^{2}\right)-F\left(m_{H^{0}}^{2},m_{A^{0}}^{2}\right)\right\} ,\nonumber \\
\varDelta S & = & \frac{1}{24\pi}\left\{ \left(2s_{\mathrm{w}}^{2}-1\right)^{2}G\left(m_{H^{\pm}}^{2},m_{H^{\pm}}^{2},M_{Z}^{2}\right)+G\left(m_{H^{0}}^{2},m_{A^{0}}^{2},M_{Z}^{2}\right)+\ln\left(\tfrac{m_{H^{0}}^{2}m_{A^{0}}^{2}}{m_{H^{\pm}}^{4}}\right)\right\} ,\hspace{0.5cm}
\end{eqnarray}
where $s_{\mathrm{W}}\equiv\sin~\theta_{W}$, with $\theta_{W}$ is
the Weinberg mixing angle, and the functions $F$ and $G$ are loop
integrals that are given in the literature~\cite{Grimus:2008nb}. 
\item \textbf{The ratio $R_{\gamma\gamma}$}: The existence of the charged
scalar $H^{\pm}$ modifies the value of the branching ratio $\mathcal{B}(h\rightarrow\gamma\gamma)$,
which both ATLAS and CMS collaborations have reported their combined
results on the ratio $R_{\gamma\gamma}:=\mathcal{B}(h\to\gamma\gamma)/\mathcal{B}(h\to\gamma\gamma)^{SM}=1.09\pm0.12$~\cite{ATLAS:2017ovn}.
In the model we are considering, $R_{\gamma\gamma}$ reads 
\begin{eqnarray}
R_{\gamma\gamma}=\left\vert 1+\frac{\lambda_{3}\upsilon^{2}}{2m_{H^{\pm}}^{2}}\frac{A_{0}^{\gamma\gamma}\left(\frac{m_{h}^{2}}{4m_{H^{\pm}}^{2}}\right)}{A_{1}^{\gamma\gamma}\left(\frac{m_{h}^{2}}{4M_{W}^{2}}\right)+N_{c}Q_{t}^{2}A_{1/2}^{\gamma\gamma}\left(\frac{m_{h}^{2}}{4m_{t}^{2}}\right)}\right\vert ^{2},
\end{eqnarray}
where the functions $A_{i}^{\gamma\gamma}$ are given in~\cite{Chen:2013vi}.
Another Higgs decay branching ratio that gets modified is $h\rightarrow\gamma Z$,
with the signal strength is given by 
\begin{eqnarray}
R_{\gamma Z}=\left\vert 1-\frac{1-2s_{\mathrm{w}}^{2}}{c_{\mathrm{w}}}\frac{\lambda_{3}\upsilon^{2}}{2m_{H^{\pm}}^{2}}\frac{A_{0}^{\gamma Z}\left(\frac{m_{h}^{2}}{4m_{H^{\pm}}^{2}},\frac{M_{Z}^{2}}{4m_{H^{\pm}}^{2}}\right)}{c_{\mathrm{w}}A_{1}^{\gamma Z}\left(\frac{m_{h}^{2}}{4M_{W}^{2}},\frac{M_{Z}^{2}}{4M_{W}^{2}}\right)+\frac{6-16s_{\mathrm{w}}^{2}}{3c_{\mathrm{w}}}A_{1/2}^{\gamma Z}\left(\frac{m_{h}^{2}}{4m_{t}^{2}},\frac{M_{Z}^{2}}{4m_{t}^{2}}\right)}\right\vert ^{2},\label{Rgz}
\end{eqnarray}
where $c_{\mathrm{w}}^{2}=1-s_{\mathrm{w}}^{2}$ and the functions
$A_{i}^{\gamma Z}$ are given in~\cite{Chen:2013vi}. The branching
ratio $h\rightarrow\gamma Z$ is not measured yet, but when it will
be measured with a good precision, it can give a hint about the extra
charged scalar whether it is a singlet or it belongs to a higher order
multiplet. 
\item \textbf{LEP direct searches of charginos and neutralinos}: The search
for the inert particles have never been done at colliders. However,
their signature is very similar to those of neutralinos and charginos
in supersymmetric models~\cite{Abdallah:2003xe}. We take a conservative
approach and impose the following lower bounds 
\begin{eqnarray}
m_{H^{\pm}}>113.5~\text{GeV}~,\quad\max\{m_{H^{0}},m_{A^{0}}\}>100~\text{GeV},
\end{eqnarray}
where the last bound comes from a re-interpretation of neutralino
searches at LEP~\cite{Lundstrom:2008ai} in the context of the IHDM. 
\item \textbf{Dark matter relic density}: In this model, we are considering
the lightest right handed neutrino to be the DM candidate. Its annihilation
occurs onto SM neutrinos and charged leptons via t-channel diagrams
mediated by the members of the Inert doublet. After computing the
thermally averaged cross section~\cite{Gondolo:1990dk}, $<\sigma_{N_{1}N_{1}}v_{r}>=<\sigma(N_{1}N_{1}\rightarrow\ell^{+}\ell^{-},\nu\bar{\nu})v_{r}>$,
the DM relic abundance can be expressed as~\cite{Jungman:1995df}
\begin{align}
\Omega_{N_{1}}h^{2}\simeq\frac{3\times10^{-26}~\text{cm}^{3}s^{-1}}{<\sigma_{N_{1}N_{1}}v_{r}>},\label{om}
\end{align}
which we require to be in agreement with the measured values by WMAP~\cite{Hinshaw:2012aka}
and Planck~\cite{Ade:2015xua} collaborations. The expression of
the relic density in (\ref{om}) is estimated without taking into
account the co-annihilation effect. Such effect can be important when
$N_{2}$ and/or $N_{3}$ have masses very close to that of $N_{1}$,
i.e., $\Delta_{i}=(M_{i}-M_{1})/M_{1}\ll1$, and hence one should
not neglect it when computing the relic density. However, co-annihilation
with the inert members, such as the process $N_{1}H^{\pm}\rightarrow\ell_{\alpha}\gamma$,
is less important due to the smallness the electromagnetic coupling
as compared to the Yukawa couplings $h_{ik}$.

In order to account for such effect, one substitute $<\sigma_{N_{1}N_{1}}v_{r}>$
in (\ref{om}) by the thermal average of the effective cross section~\cite{Kong:2005hn}
\begin{eqnarray}
\sigma_{eff}(x_{f}) & = & \sum_{i,k}^{3}\tfrac{2^{2}}{g_{eff}^{2}}\left(1+\Delta_{i}\right)^{3/2}\left(1+\Delta_{k}\right)^{3/2}e^{-x_{f}\left(\Delta_{i}+\Delta_{k}\right)}\left\langle \sigma_{ik}v_{r}\right\rangle %,
\end{eqnarray}
with 
\begin{eqnarray}
g_{eff}(x_{f}) & = & \sum_{i}^{3}2\left(1+\Delta_{i}\right)^{3/2}e^{-x_{f}\Delta_{i}},\label{sigeff}
\end{eqnarray}
Here, $x_{f}=M_{1}/T_{f}$ is the freeze-out parameter, $g_{eff}(x_{f})$
accounts for the effective multiplicity at the freeze-out, and $\left\langle \sigma_{ik}v_{r}\right\rangle $
is the thermally averaged cross section of $N_{i}N_{k}\rightarrow\ell_{\alpha}^{-}\ell_{\beta}^{+},\nu_{\alpha}\bar{\nu}_{\beta}$
at the freeze-out. Here, the ratio 
\begin{eqnarray}
\Delta_{i}=\frac{M_{i}-M_{1}}{M_{1}},\label{eq:Di}
\end{eqnarray}
is the relative mass difference. The cross section formulas for the
processes $N_{i}N_{k}\rightarrow\ell_{\alpha}^{-}\ell_{\beta}^{+},\nu_{\alpha}\bar{\nu}_{\beta}$
are given in Eq. (\ref{eq:Sig}) in the appendix. 
\item \textbf{Dark matter direct detection}: Although $N_{1}$ does not
couple directly to the Higgs boson or $Z$ gauge boson, it acquire
an effective vertex $hN_{1}\bar{N}_{1}$ at one loop. In this case,
the spin independent (SI) scattering cross section of $N_{1}$ off
a nucleon ${\cal N}$ reads 
\begin{eqnarray}
\sigma_{det}(N_{1}+{\cal N}\rightarrow N_{1}+{\cal N}) & = & \frac{\tilde{y}_{hN_{1}\bar{N}_{1}}^{2}(m_{{\cal N}}-\frac{7}{9}m_{{\cal B}})^{2}m_{{\cal N}}^{2}M_{1}^{2}}{4\pi\upsilon^{2}m_{h}^{4}(m_{{\cal N}}+M_{1})^{2}},\label{sigma}
\end{eqnarray}

where $m_{{\cal N}}$ and $m_{{\cal B}}$ are the nucleon and baryon
masses in the Chiral limit~\cite{He:2008qm}, and $\tilde{y}_{hN_{1}\bar{N}_{1}}$
is the induced one loop effective dark matter coupling to the Higgs
boson. There are three generic contributions to the effective $\tilde{y}_{hN_{1}\bar{N}_{1}}$
coupling that lead to the SI interaction as are shown in Fig.~\ref{fig:hNN}.
\begin{figure}[h]
\centering \includegraphics[width=0.95\linewidth]{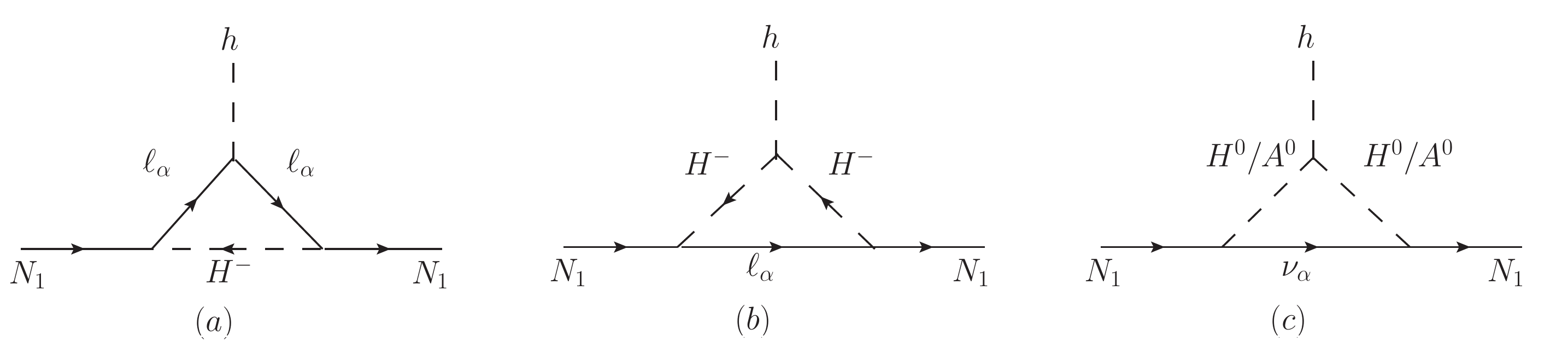} \caption{Feynman diagrams that are responsible to the effective coupling $\tilde{y}_{hN_{1}\bar{N}_{1}}$.}
\label{fig:hNN} 
\end{figure}

In non-relativistic limit and $m_{\ell_{\alpha}}\ll M_{1}$, the effective
coupling can be approximated by~\cite{Okada:2013rha}\footnote{The exact formula is derived in Appendix~\ref{appendix2}.}
\begin{eqnarray}
\tilde{y}_{hN_{1}N_{1}} & = & -\frac{\lambda_{3}\upsilon\sum_{\alpha}|h_{\alpha1}|^{2}}{16\pi^{2}M_{1}^{3}}\left(M_{1}^{2}+\left(m_{H^{\pm}}^{2}-M_{1}^{2}\right)ln\left(\frac{m_{H^{\pm}}^{2}-M_{1}^{2}}{m_{H^{\pm}}^{2}}\right)\right)\nonumber \\
 & & -\frac{\left(\lambda_{3}+\lambda_{4}/2\right)\upsilon\sum_{\alpha}|h_{\alpha1}|^{2}}{16\pi^{2}M_{1}^{3}}\left(M_{1}^{2}+\left(\bar{m}^{2}-M_{1}^{2}\right)ln\left(\frac{\bar{m}^{2}-M_{1}^{2}}{\bar{m}^{2}}\right)\right),\label{yhNN}
\end{eqnarray}
where $\bar{m}^{2}=(m_{H^{0}}^{2}+m_{A^{0}}^{2})/2$. In the above
expression, the first [second] line corresponds to diagram (b),
and the second line [(c)], whereas diagram (a) contribution is
negligible since it is $m_{\ell_{\alpha}}^{2}/m_{H^{\pm}}^{2}$ proportional.
Thus, in our scan of the parameter space we will impose the recent
bounds on the dark matter-nucleon scattering cross section from the
LUX~\cite{Akerib:2016vxi} and XENON1T~\cite{Aprile:2017iyp} direct
detection experiments. 
\end{itemize}

\section{Numerical Analysis and Discussion}

\label{sec:analysis}

The model contains 26 free parameters: 4 quartic couplings in the
scalar potential $\lambda_{i},i=2..5$, 18 (9 complex) Yukawa couplings
$h_{ij},i,j=1..3$, 3 RHNs masses $M_{k},k=1..3$ and the squared-mass
parameter $\mu_{2}^{2}$ of the inert Higgs. In our numerical analysis,
we perform a scan of those parameters over the parameters ranges 
\begin{eqnarray}
0<\lambda_{2},|\lambda_{3,4}|<2,\nonumber \\
1\textrm{ GeV}<M_{1}<M_{2}<M_{3}<2\textrm{ TeV},\nonumber \\
-\upsilon^{2}<\mu_{2}^{2}<7\upsilon^{2},\qquad10^{-5}<|h_{ij}|<1.\label{scan}
\end{eqnarray}

\begin{figure}[!h]
\centering \includegraphics[width=0.5\linewidth]{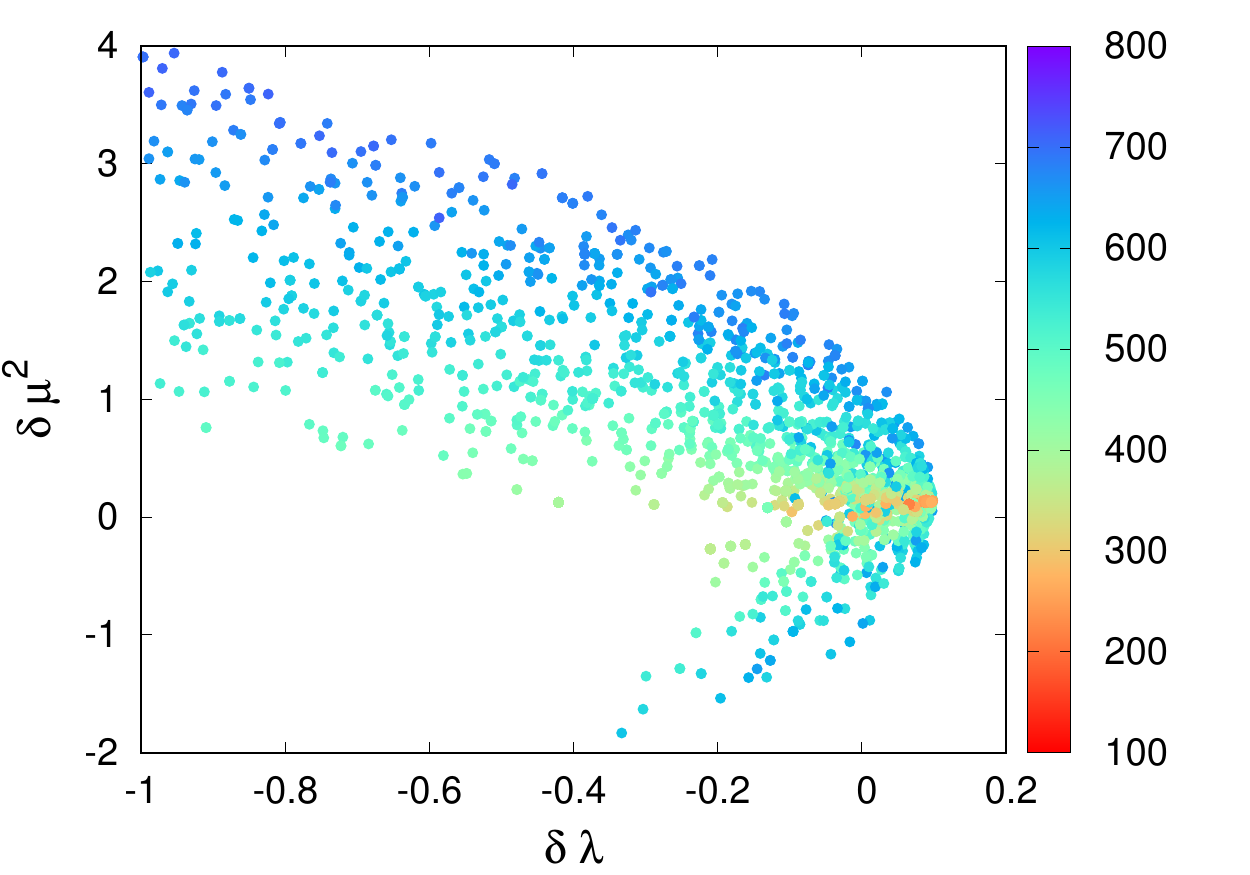}~\includegraphics[width=0.5\linewidth]{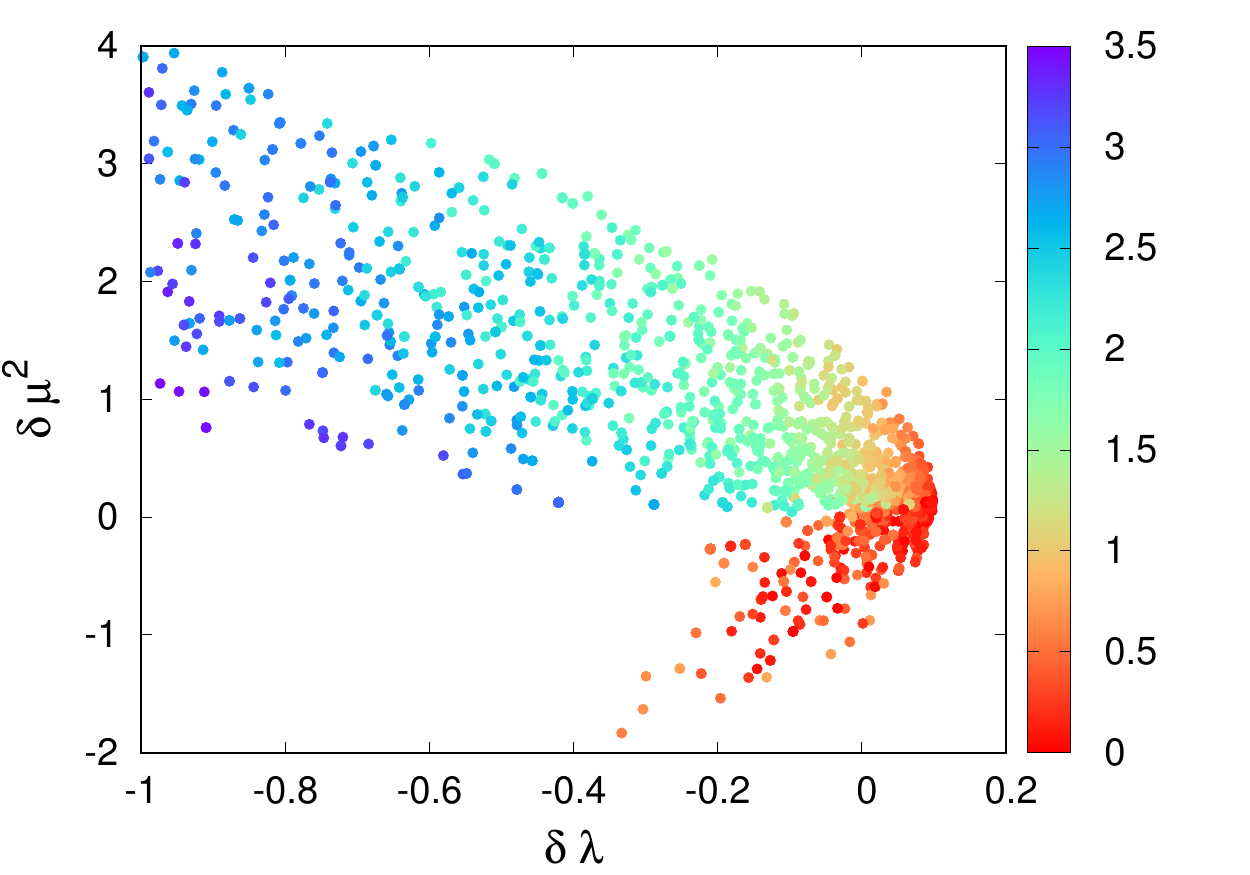}
\caption{The ratios $\delta\lambda$ and $\delta\mu^{2}$ that represent the
relative corrections due to the inert members. In the palette, we
read the large inert members mass (right) and its coupling to the
Higgs, i.e., $|\lambda_{3}|$ ($|\lambda_{3}+\lambda_{4}/2\pm\lambda_{5}/2|$)
if $m_{H^{+}}>m_{H^{0},A^{0}}$ ($m_{H^{+}}<m_{H^{0},A^{0}}$).}
\label{fig:dX} 
\end{figure}

Taking into account all theoretical and experimental constraints mentioned
in the previous section, we scan over the parameters range (\ref{scan}).
Within this parameters range, one can estimate the effect of the one-loop
corrections in (\ref{1l}) on the observables $X=\lambda,\,\mu^{2}$,
by showing the ratio $\delta X=\left(X-X_{tree-level}\right)/X_{tree-level}$
in Fig.~\ref{fig:dX}, where we consider 3000 benchmark points that
fulfill all the conditions mentioned in the previous section. 

One has to notice that the one-loop effect is very important for massive
and strongly coupled inert members. For instance, the inert corrections
in (\ref{1l}) gives $\delta\lambda=-1\sim0.098$ and $\delta\mu^{2}=-1.83\sim3.94$,
i.e., the Higgs mass and quartic coupling could be fully radiative
$\delta\lambda\sim-1$.

\begin{figure}[!h]
\centering \includegraphics[width=0.5\linewidth]{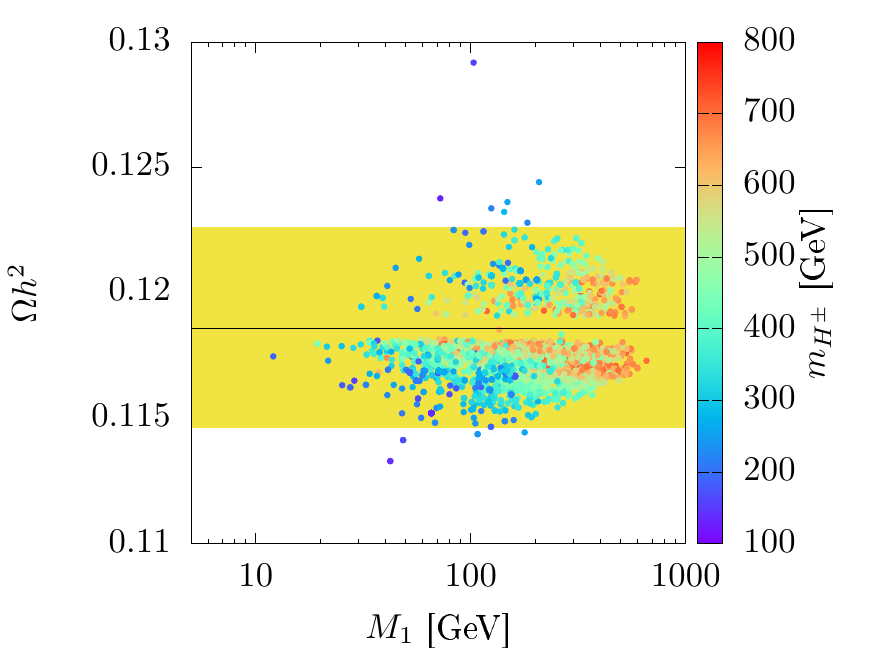}~\includegraphics[width=0.5\linewidth]{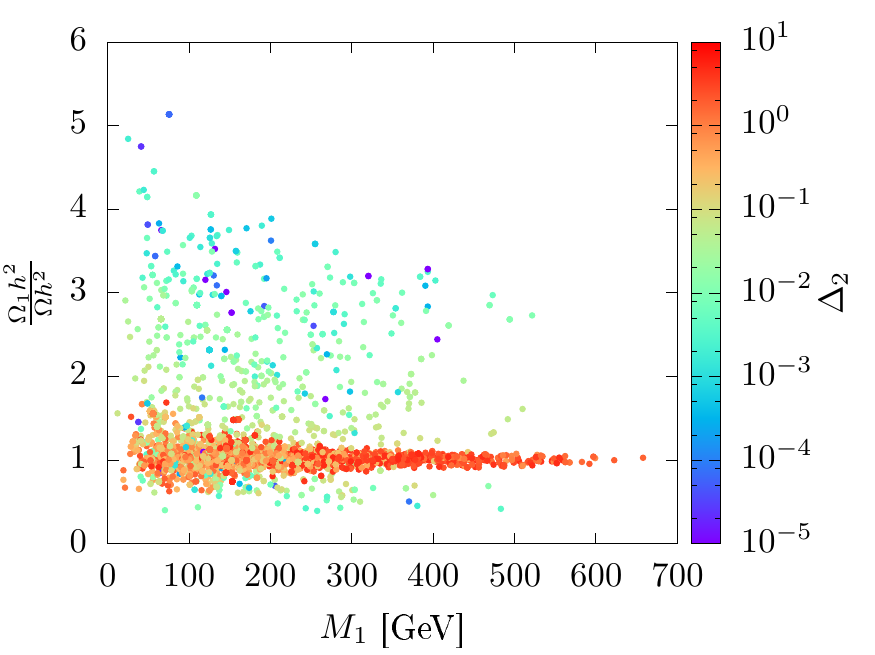}\\
 \includegraphics[width=0.5\linewidth]{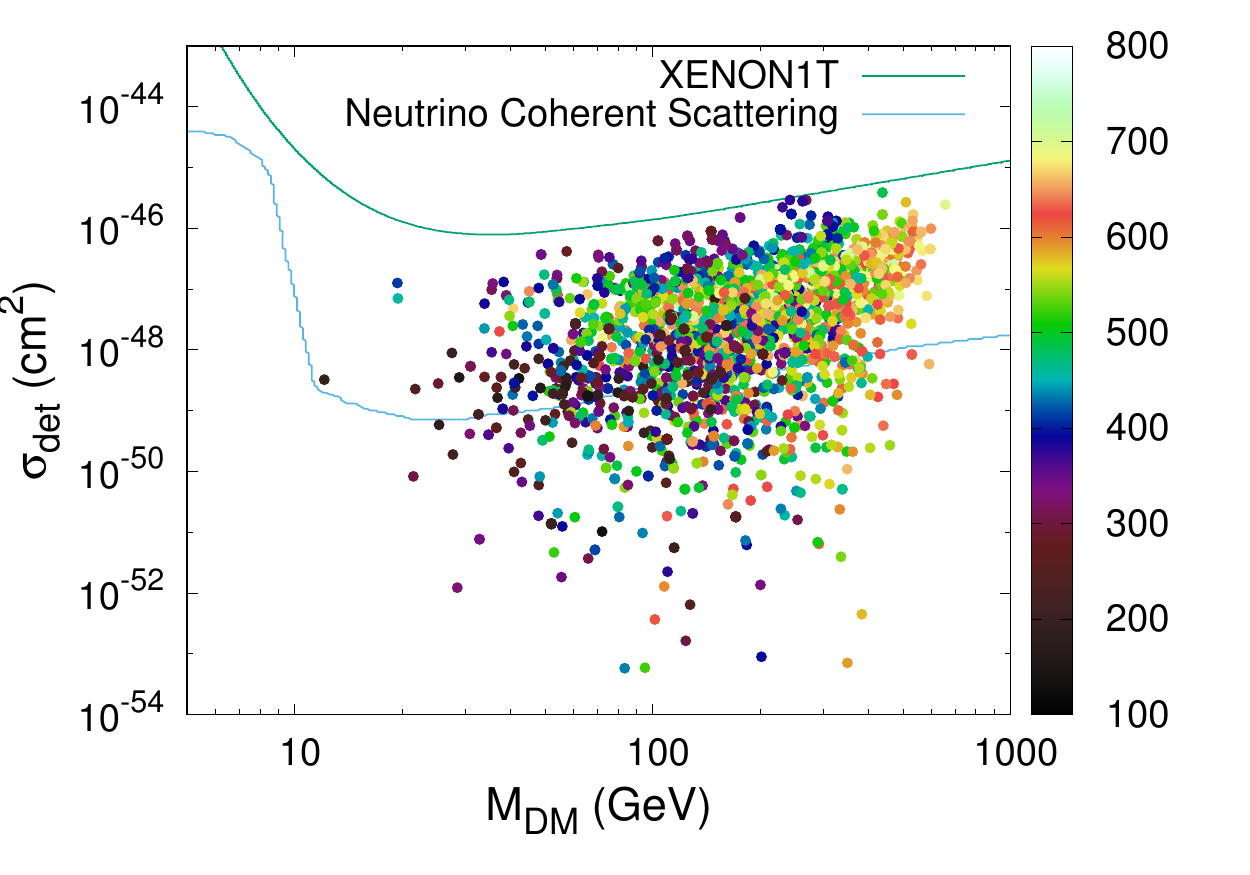} \caption{Top left: the relic density of the DM as a function of its mass. The
yellow band shows the experimental measurement of $\Omega_{N_{1}}h^{2}$
at $2\sigma$. Top right: the ratio of $\Omega_{1}h^{2}/\Omega h^{2}$,
where $\Omega_{1}h^{2}$ ($\Omega h^{2}$) does (not) include the
co-annihilation effect. The palette represents the degeneracy parameter
$\Delta_{2}$ defined in (\ref{eq:Di}). Bottom: the direct detection
cross section as a function of the DM mass, with the exclusion plot
obtained from the Xenon1T experiment. The blue curve represents the
irreducible neutrino background due to the coherent neutrino-nucleus
scattering.}
\label{fig:DM} 
\end{figure}

In Fig.~\ref{fig:DM}, we depict the DM relic density, co-annihilation
effect and the direct detection spin-independent cross section as
a function of the DM mass for the benchmark points used in Fig.~\ref{fig:dX}.

From Fig.~\ref{fig:DM}, one remarks that: (1) most of the benchmark
points are within the $1\sigma$ envelope on the relic density measurement.
(2) We see that the lightest RHN can be a viable DM candidate for
masses in the range $10-700~\textrm{GeV}$ and with a spin-independent
direct detection cross section below the experimental bound. Note
that a large fraction of the benchmark points are above the irreducible
neutrino background and hence can be probed in future direct detection
experiments. (3) The ratio co-annihilation effect that is presented
with the ratio $\Omega_{1}h^{2}/\Omega h^{2}$, where $\Omega_{1}h^{2}$
($\Omega h^{2}$) refers to the relic density within (without) the
co-annihilation effect. As expected, the co-annihilation effect is
important only for benchmark points where the two RHN's ($N_{1},N_{2}$)
are degenerate, i.e., very small mass difference values. It is worth
mentioning that even though one could allow $M_{1}$ to vary over
the mass range in (\ref{scan}), for DM heavier than $700~GeV$ the
relic density requires the couplings $h_{i,1}$ to be much larger
than unity so that the relic density will be in agreement with the
observed value. However, this will make the constraints on the LFV
processes very difficult to satisfy.

\begin{figure}[!h]
\centering \includegraphics[width=0.5\textwidth]{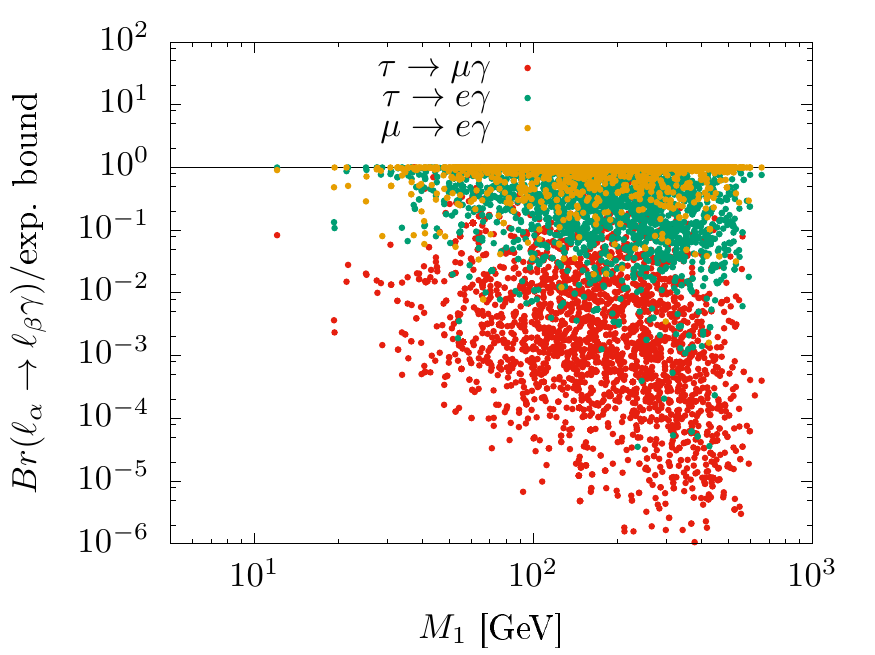}~\includegraphics[width=0.5\textwidth]{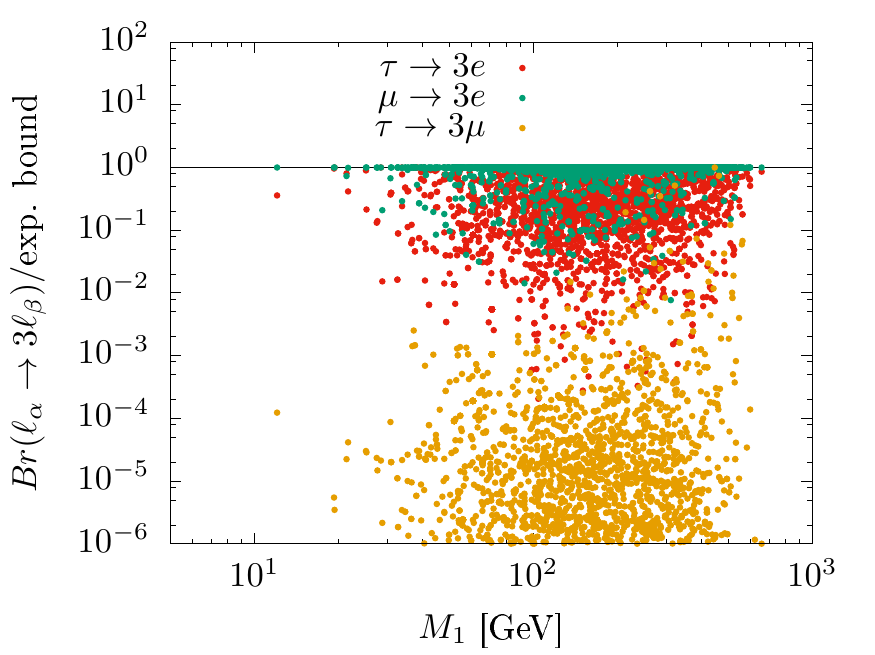}
\caption{The branching ratios $\text{Br}(\ell_{\alpha}\to\ell_{\beta}+\gamma)$
(left) and $\text{Br}(\ell_{\alpha}\to3\ell_{\beta})$ (right) normalized
by their experimental bounds as a function of $M_{1}$}
\label{fig:LFV} 
\end{figure}

For the same benchmark points, we present in Fig.~\ref{fig:LFV},
the different branching fractions of the LFV processes $\ell_{\alpha}\to\ell_{\beta}+\gamma$
(left panel) and $\ell_{\alpha}\to3\ell_{\beta}$ (right panel), normalized
to their experimental bounds, versus the DM mass. We Note that the
stringent lepton flavor violation constraint comes from the process
$\mu\to e\gamma$, where it is severely fulfilled for most of the
benchmark points. This could be achieved by cancellation between different
terms in (\ref{meg}), since the right relic density value does not
allow the Yukawa couplings to have smaller values.

\begin{figure}[!h]
\centering \includegraphics[width=0.5\textwidth]{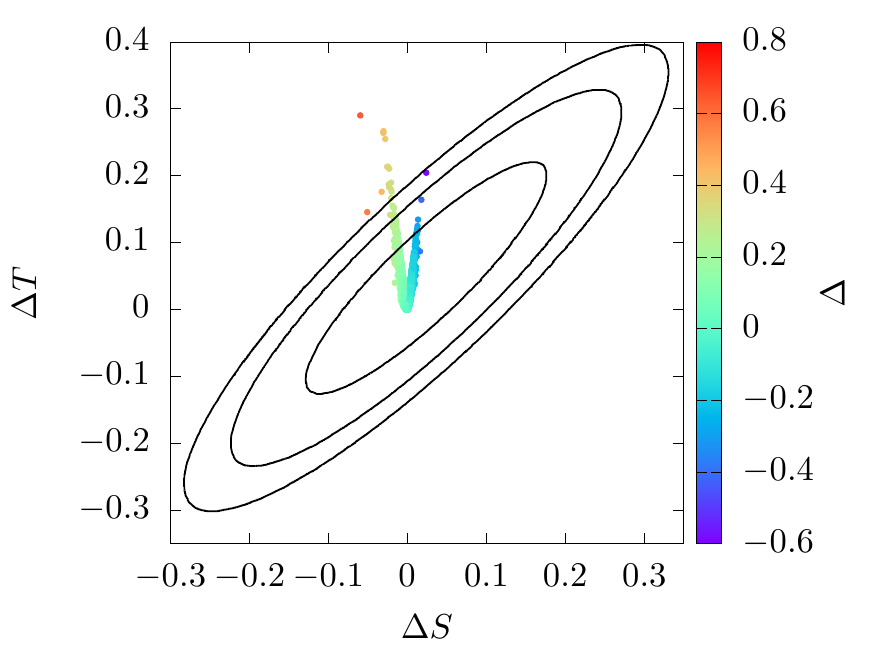}~\includegraphics[width=0.5\textwidth]{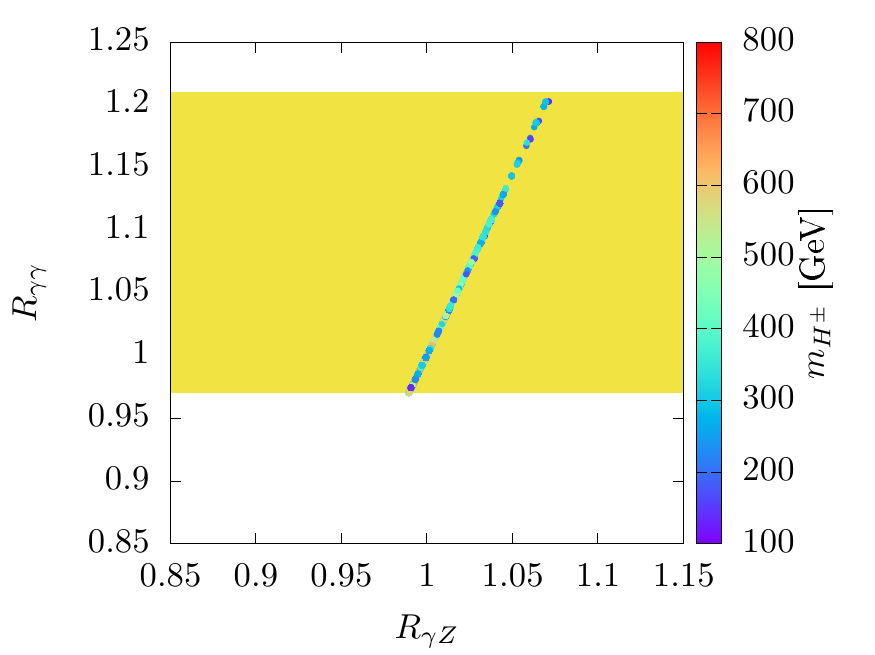}
\caption{Left: Constraints due to oblique parameters $S$ and $T$ with the
palette showing the ratio $\Delta$ defined in (\ref{Delta}). Right:
the scatter plot of ($R_{\gamma\gamma}$,$R_{\gamma Z}$), where the
palette shows the value of $m_{H^{\pm}}$ allowed by theoretical and
experimental constraints. The yellow band in the right panel shows
the $1\sigma$ interval of $R_{\gamma\gamma}$ of both ATLAS and CMS.}
\label{fig:Constr} 
\end{figure}

In Fig.~\ref{fig:Constr}-left, we depict the oblique parameters
$S$ versus $T$ where the different ellipses represent the $68\%$,
$95\%$ and $99\%$ CL intervals obtained from the precise measurements
of various observables. The color map shows a ratio defined by

\begin{eqnarray}
\Delta=\frac{2m_{H^{\pm}}-m_{A^{0}}-m_{H^{0}}}{2m_{H^{\pm}}},\label{Delta}
\end{eqnarray}
which parametrizes how close is the charged Higgs mass to the arithmetic
mean of the CP-even and CP-odd Higgs masses. In the right panel of
Fig.~\ref{fig:Constr}, we show $R_{\gamma\gamma}$ vs $R_{\gamma Z}$.
As can be seen from left panel in the figure, one remarks that the
electroweak precision tests exclude the benchmark points with larger
values of the parameter $\Delta$. From Fig.~\ref{fig:Constr}-right,
one can see that the two branching ratios are proportional to each
other due to the nature of the new additional scalar multiplet, i.e.,
the inert doublet. This can be understood from the factor $(1-2s_{\mathrm{w}}^{2})/c_{\mathrm{w}}$
in (\ref{Rgz}).

\begin{figure}[!h]
\centering \includegraphics[width=0.5\textwidth]{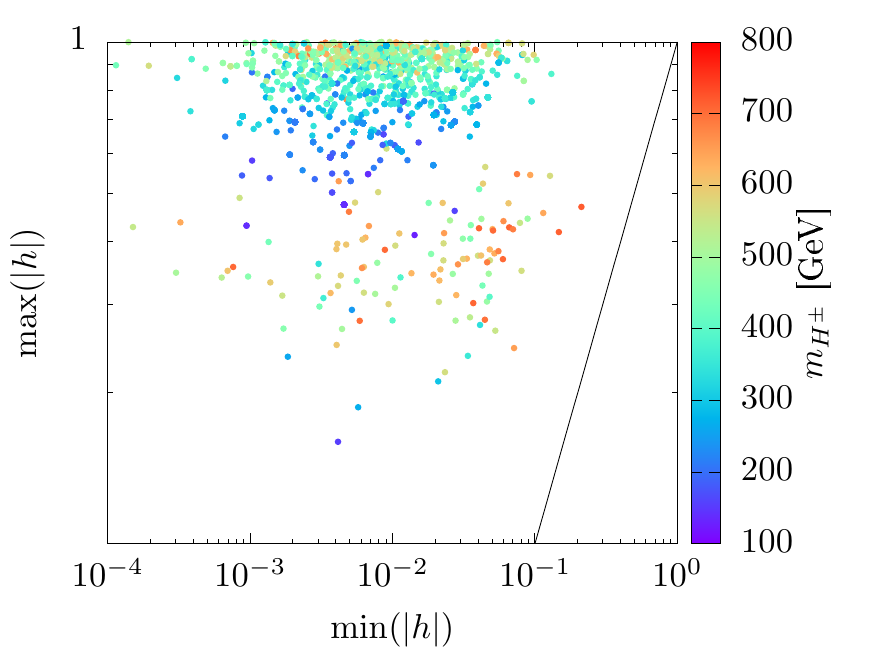}~\includegraphics[width=0.5\textwidth]{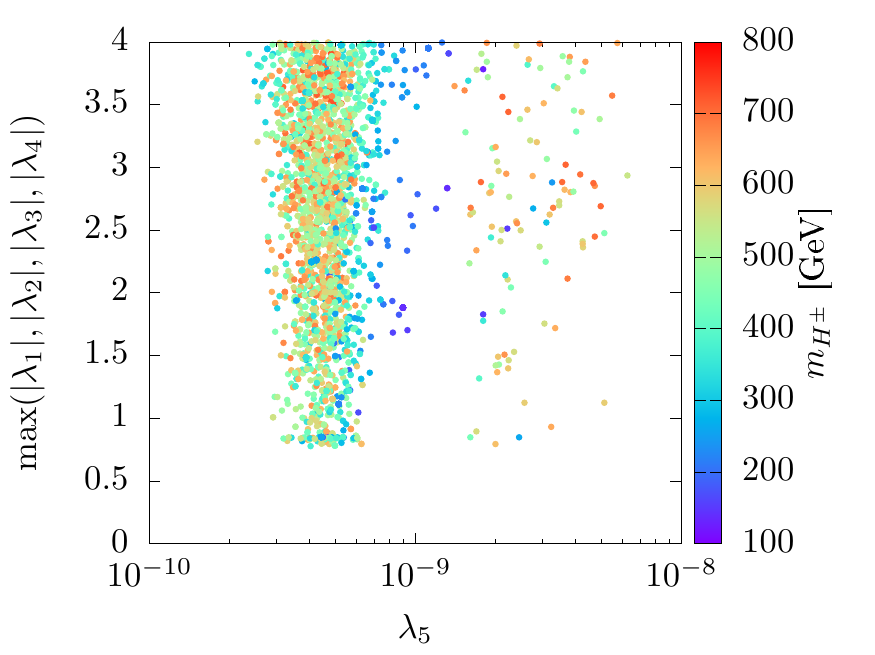}\\
 \includegraphics[width=0.5\textwidth]{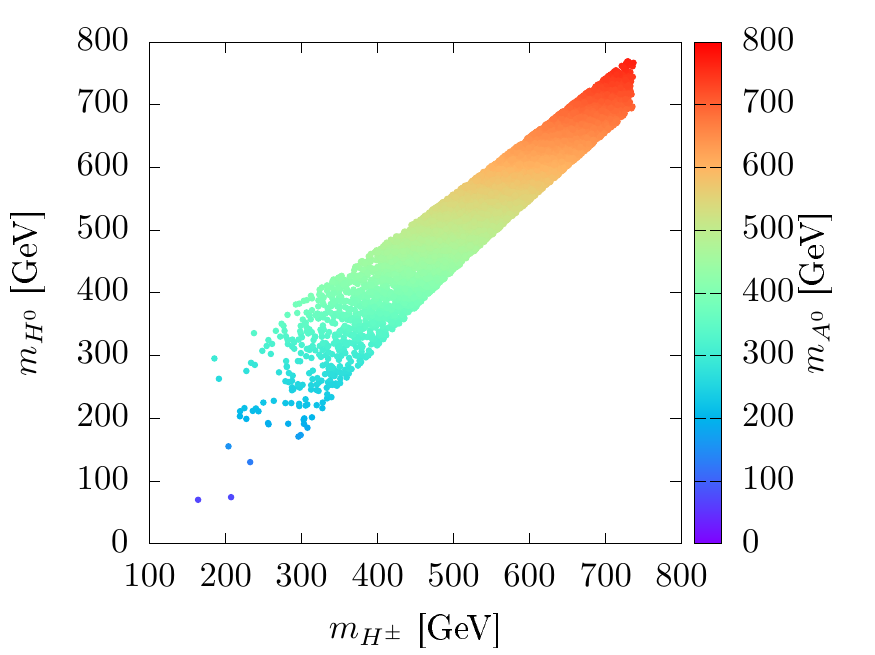} \caption{Allowed model parameters by various constraints. Top left: $\max(|h|)$
versus $\min(|h|)$ where $h\equiv h_{ij}$ are the new Yukawa couplings
(\ref{LL}). Top right: $\max(\lambda_{1},\lambda_{2},|\lambda_{3}|,|\lambda_{4}|)$
versus $\lambda_{5}$. Bottom panel: scatter plot of $m_{H^{\pm}}$
versus $m_{H^{0}}$, where $m_{A^{0}}$ is shown in the palette.}
\label{fig:parameters} 
\end{figure}

We give in Fig.~\ref{fig:parameters} the range of parameter space
of the model after passing all the theoretical and experimental constraints,
leading to the results presented in the previous figures. In the left
top panel, we show $\min(|h|)$ versus $\max(|h|)$ where $h\equiv h_{ij}$
are the new Yukawa couplings. The line displayed in that figure define
the regime couplings are degenerate couplings regime, and farther
from it the less degenerate the couplings are. Moreover, the Yukawa
h-couplings values that fit the neutrino oscillation, fulfill the
LFV and the relic density have the same order of magnitude whether
the CP violating phases are vanishing or not. Thus, in this model
there is no favored values or range of CP violating phases. Furthermore,
in both cases of normal and inverted neutrino mass hierarchy, the
allowed parameter space is similar. For the case of real Yukawa h-couplings,
they need to be about one order of magnitude smaller than the complex
case of complex, which is due to the fact that the cancellations in
(\ref{meg}) required to fulfill the LFV constraints is much easier
to achieve for complex valued Yukawa couplings than real ones.

In the right top panel of the same figure, we show $\max(|\lambda_{1,2,3,4}|)$
as a function of the coupling $|\lambda_{5}|$. One can see that $\lambda_{5}$
take very small values, and this is required to be in agreement with
the observed neutrino masses and mixing angles. This allow the new
Yukawa couplings $h_{ij}$ to be unsuppressed as opposed to the case
considered in~\cite{Ma:2006km}. The bottom panel, represents the
allowed masses of the CP-odd scalar, $m_{A^{0}}$, projected on the
$(m_{H^{0}},m_{H^{\pm}})$ plan. The implication of the smallness
of $\lambda_{5}$ is that the CP-odd and CP-even states are quasi
degenerate, which render the constraints on the oblique parameters
easy to satisfy.

\section{Collider phenomenology}

\label{sec:collider}

\subsection{Possible signatures}

The RHNs, $N_{i}$, can be pair produced through several processes
both at hadron as well at lepton colliders. At hadron colliders, however,
it cannot be produced directly because of the absence of the vertices
$Z^{0}N\bar{N}$, $\gamma N\bar{N}$ and $hN\bar{N}$ in this model,
and can only be found in the decay products of the inert doublet members.
At lepton colliders such as the ILC or the FCC-ee, $N_{i}$ can be
produced in $t$-channel processes with charged boson exchange. For
the lightest RHN, $N_{1}$, when produced at colliders, it is accounted
as missing energy, however, $N_{2}$ and $N_{3}$ depending on their
decay width $\Gamma(N_{2,3}\rightarrow N_{1}+\ell^{+}\ell^{-})$ may
or may not behave as missing energy. This depends on the decay lengths
$l_{i}=E_{i}/M_{i}/\Gamma_{N_{i}}$ (i=2,3), with $E_{i}$ and $\Gamma_{N_{i}}$
are the energy and total decay width of $N_{i}$, respectively.

\begin{figure}[!h]
\centering \includegraphics[width=0.5\textwidth]{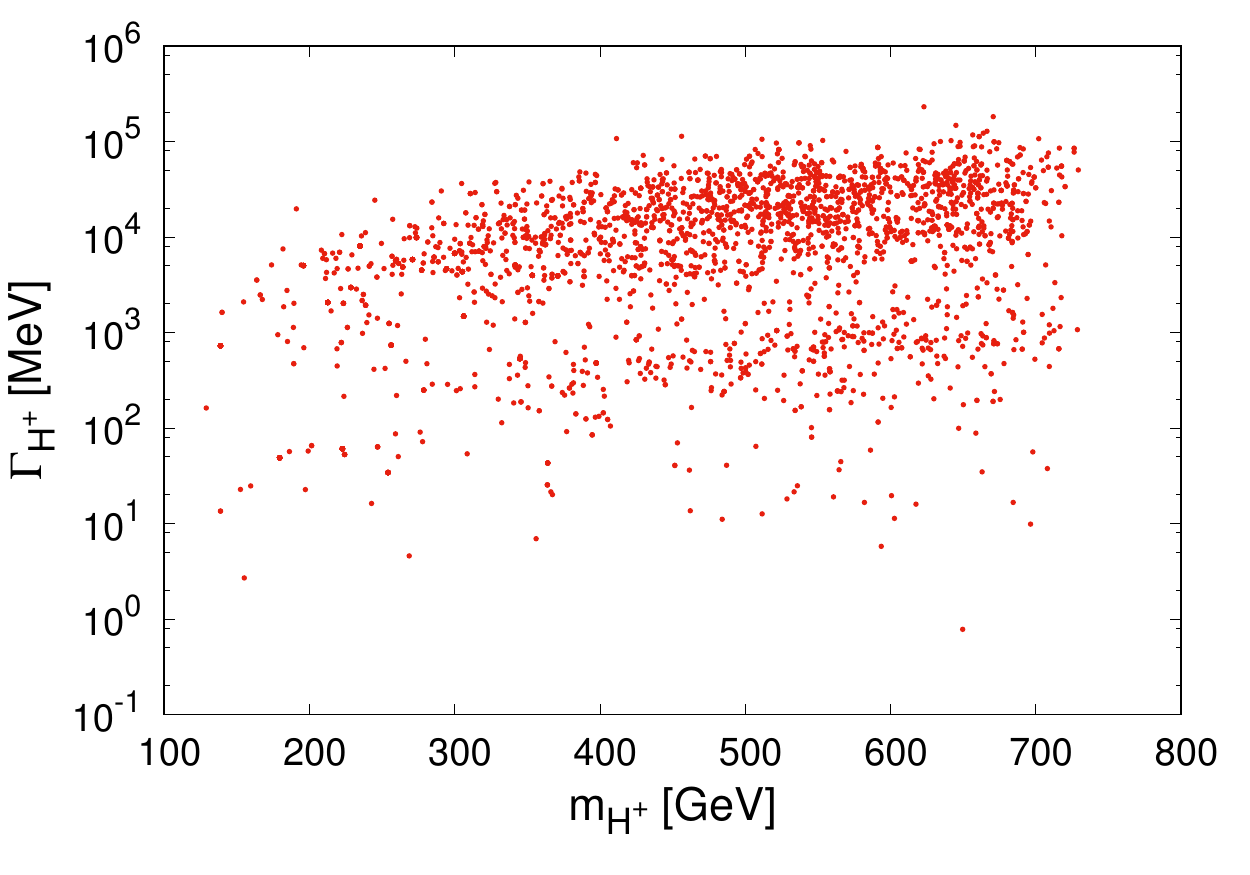}~\includegraphics[width=0.5\textwidth]{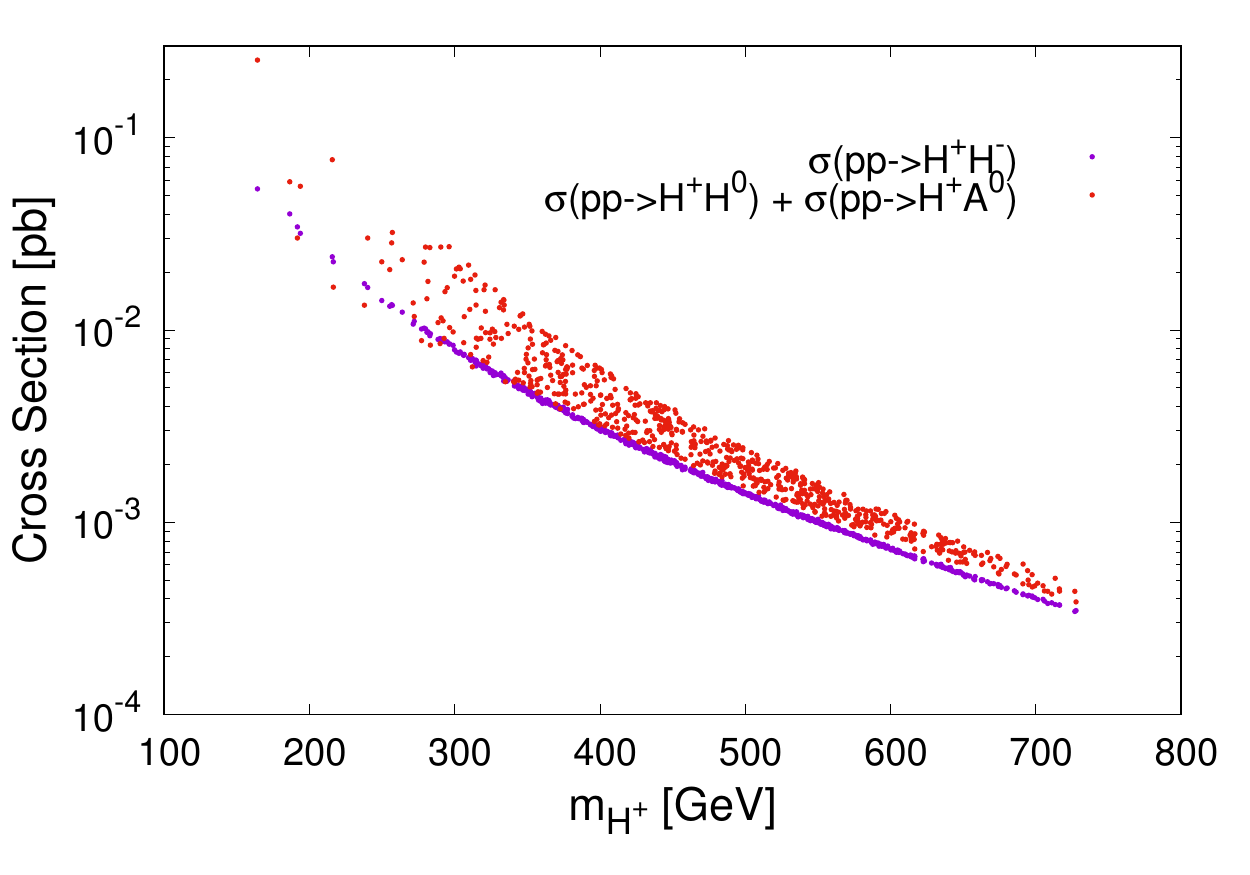}
\caption{Total width of the charged Higgs boson (left panel) and its inclusive
cross section (right panel) as a function of of $m_{H^{\pm}}$.}
\label{fig:charged} 
\end{figure}

From the different theoretical and experimental constraints, especially
from neutrino mass and the electroweak precision observables, the
neutral scalar particles are degenerate and hence the decays such
as $A^{0}\to H^{0}Z^{0}$ are kinematically forbidden. Thus, these
particles can only undergo invisible decays (if kinematically allowed)
into a RHN $N_{i}$ and a SM neutrino. At the LHC, the charged Higgs
boson can be produced in pair or in association with $A^{0}/H^{0}$
with $\sigma\simeq200$ fb for light charged Higgs boson mass in the
$H^{0}H^{\pm}(A^{0}H^{\pm})$ mode (see Fig.~\ref{fig:charged}).
This channel, for $\Delta_{H^{\pm}}=m_{H^{\pm}}-m_{H^{0},A^{0}}<M_{W^{\pm}}$,
leads exclusively to the spectacular mono-lepton signature. While
dilepton signals can be observed in the case of the pair production
of charged Higgs boson again for $\Delta_{H^{\pm}}<M_{W^{\pm}}$.
On the other hand, for $\Delta_{H^{\pm}}>M_{W^{\pm}}$, the decays
$H^{\pm}\to A^{0}/H^{0}W^{\pm}$ are kinematically allowed and hence
a wide range of signatures are possible (see Table~\ref{table:signature}).
For the case of $H^{0}H^{0}$ ($A^{0}A^{0}$ or $H^{0}A^{0}$) channels,
the two particles decay exclusively into a $N_{i}\nu_{\ell}$ giving
a mono-jet or mono-photon signature where the additional jet/photon
is produced from Initial State Radiation (ISR) off the scattered quarks.
However, due to the large $\alpha_{s}(Q)/\alpha_{\textrm{em}}(Q)$
ratio, QCD radiation is dominant and mono-jet signature is more relevant
in this case.

At lepton colliders, the $N_{1}$ can be pair produced through $H^{\pm}$
exchange leading to mono-photon signature where the additional photon
is emitted either from the $e^{\pm}$ lines or from the intermediate
charged Higgs boson. Pair production of charged Higgs boson or a CP-odd
(CP-even) particle is also possible at lepton colliders. In table~\ref{table:signature},
we summarize the different signatures that can be used to either look
for $N_{1}$ in the next LHC-run or to constrain the model using the
old measurements.

In Fig.~\ref{fig:charged}, we show the total decay width of the
charged Higgs (left) as a function of its mass for some of the benchmarks
used previously. We see that the width can be as large as few hundred
of GeV for heavy charged Higgs bosons. The reason is that scalar decays
$H^{\pm}\to W^{\pm}A^{0}/H^{0}$ are kinematically allowed in certain
regions of the parameter space. In the right panel, we present the
cross section at the LHC at $\sqrt{s}=14$ TeV of $H^{\pm}H^{\mp}$
and $H^{\pm}H^{0}(A^{0})$ as a function of the charged Higgs mass.
We can see that the $H^{\pm}H^{0}(A^{0})$ have larger cross sections
and approache $200$ fb for light charged Higgs boson. We stress that
for heavy scalar masses, the production cross section is extremely
small. This makes the observation of the new states extremely difficult
at moderate luminosities.

\begin{table}[h]
\begin{centering}
\begin{adjustbox}{max width=\textwidth} %
\begin{tabular}{c|c|c}
\hline 
Process & Decay mode & signature \tabularnewline
\hline 
\hline 
$pp,e^{+}e^{-}\to H^{\pm}H^{\mp}$ & $H^{\pm}\to W^{\pm}H^{0}/A^{0}\to q_{1}\bar{q}_{2}N_{1}\nu_{\ell}$,
$H^{\mp}\to W^{\mp}H^{0}/A^{0}\to q_{3}\bar{q}_{4}N_{1}\bar{\nu}_{\ell}$ & $4\textrm{ jets}+\slashed{E}_{T}$ \tabularnewline
\hline 
 & $H^{\pm}\to W^{\pm}H^{0}/A^{0}\to\ell^{\pm}\nu_{\ell}N_{1}\nu_{\ell}$,
$H^{\mp}\to W^{\mp}H^{0}/A^{0}\to q_{3}\bar{q}_{4}N_{1}\bar{\nu}_{\ell}$ & $1\ell+2\textrm{ jets}+\slashed{E}_{T}$ \tabularnewline
\hline 
 & $H^{\pm}\to W^{\pm}H^{0}/A^{0}\to\ell_{1}^{\pm}\nu_{\ell}N_{1}\nu_{\ell}$,
$H^{\mp}\to W^{\mp}H^{0}/A^{0}\to\ell_{2}^{\mp}\bar{\nu}_{\ell}N_{1}\nu_{\ell}$ & $2\ell+\slashed{E}_{T}$ \tabularnewline
\hline 
 & $H^{\pm}\to W^{\pm}H^{0}/A^{0}\to\ell_{1}^{\pm}\nu_{\ell}N_{1}\nu_{\ell}$,
$H^{\mp}\to\ell_{2}^{\mp}N_{1}$ & $2\ell+\slashed{E}_{T}$ \tabularnewline
\hline 
 & $H^{\pm}\to W^{\pm}H^{0}/A^{0}\to q_{1}\bar{q}_{2}N_{1}\nu_{\ell}$,
$H^{\mp}\to\ell^{\mp}N_{1}$ & $1\ell+2\textrm{ jets}+\slashed{E}_{T}$ \tabularnewline
\hline 
\hline 
$pp,e^{+}e^{-}\to H^{\pm}H^{0}/A^{0}$ & $H^{\pm}\to W^{\pm}H^{0}/A^{0}\to q_{1}\bar{q}_{2}N_{1}\nu_{\ell}$,
$H^{0}\to N_{1}\nu_{\ell}$ & $1\ell+2\textrm{ jets}+\slashed{E}_{T}$ \tabularnewline
\hline 
 & $H^{\pm}\to W^{\pm}H^{0}/A^{0}\to\ell^{\pm}\nu_{\ell}N_{1}\nu_{\ell}$,
$H^{0}\to N_{1}\nu_{\ell}$ & $1\ell+\slashed{E}_{T}$ \tabularnewline
\hline 
 & $H^{\pm}\to N_{1}\ell^{\pm}$ & $1\ell+\slashed{E}_{T}$ \tabularnewline
\hline 
\hline 
$pp,e^{+}e^{-}\to H^{0}H^{0}$ & $H^{0}\to N_{1}\nu_{\ell}$ & $\slashed{E}_{T}+\textrm{ ISR}$ (mono-jet, mono-$\gamma$) \tabularnewline
\hline 
\hline 
$e^{+}e^{-}\to N_{1}N_{1}\gamma$ & stable final state & $\slashed{E}_{T}+\gamma$ \tabularnewline
\hline 
\end{tabular}\end{adjustbox} 
\par\end{centering}
\caption{Summary of various signatures of the inert Higgs doublet members and
the lightest RHN at the LHC and lepton colliders.}
\label{table:signature} 
\end{table}

In the rest of this section, we will investigate the monolepton signature
at the LHC, that is mentioned in Table.~\ref{table:signature}.

\subsection{Monolepton signature at the LHC}

\begin{figure}[!h]
\centering \includegraphics[width=0.5\linewidth]{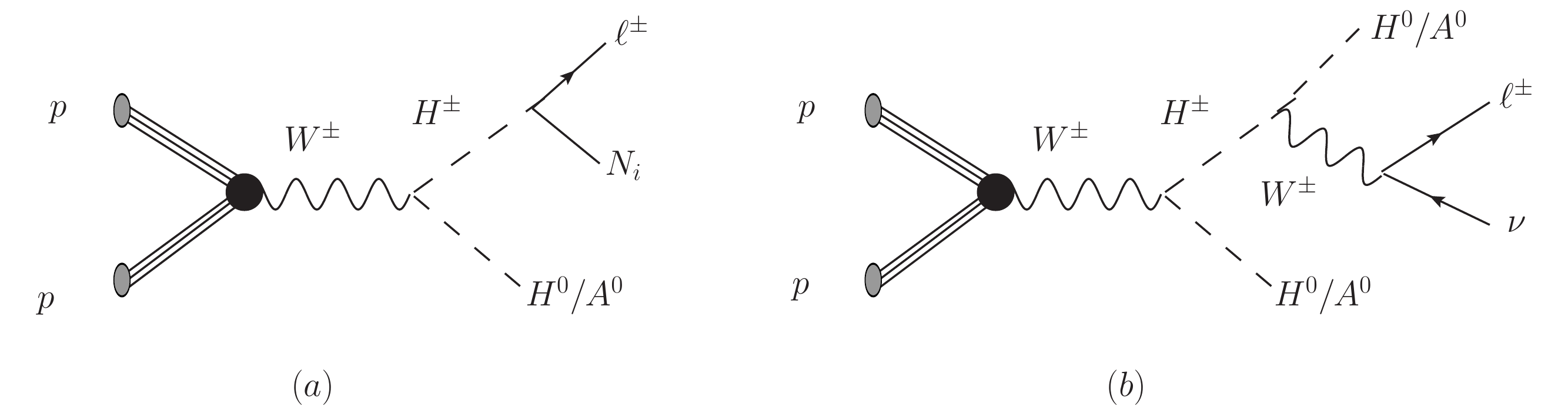} \caption{Feynman diagrams for the charged monolepton signature.}
\label{singleL} 
\end{figure}

As a benchmark study, we consider the $1\ell+E_{T}^{\textrm{miss}}$
($\ell=e,\mu$) final state in $pp$ collisions at $\sqrt{s}=14$
TeV and a luminosity of $300$ fb$^{-1}$. Such a signature can arise
from $H^{\pm}H^{0}/A^{0}$ production followed by the decay of the
charged Higgs either into 1) a charged lepton and a Majorana fermion
or 2) a $W$ gauge boson and a dark Higgs where the gauge boson decays
leptonically. Feynman diagrams showing the $1\ell+E_{T}^{\textrm{miss}}$
in the signal process are depicted in Fig.~\ref{singleL}.

Two benchmark points, denoted by \textsf{BP1} and \textsf{BP2}, are
considered in this study where the corresponding parameters are given
in Table.~\ref{benchmark.points}. Moreover, the chosen benchmarks
satisfy all the theoretical and experimental constraints and yield
a cross section of order $460$ fb and $127$ fb for \textsf{BP1}
and \textsf{BP2}, respectively. A contribution from $H^{0}H^{0}W^{\pm},A^{0}A^{0}W^{\pm}$
with the $W$-boson decaying leptonically is relevant in \textsf{BP1}
since $\sigma(pp\to H^{0}H^{0}W^{\pm}\to\ell^{\pm}E_{T}^{\mathrm{miss}})\approx45$
fb. However, the same contribution is very negligible for \textsf{BP2}
and will not be taken into account. Furthermore, they have distinct
features regarding collider phenomenology and dark matter relic density.
For the first benchmark point (\textsf{BP1}), co-annihilation effects
are important since $\Delta_{2}\approx0.08$, whereas, due to the
relatively larger value of $\Delta_{2}$, co-annihilation effects
are negligible for the second benchmark (\textsf{BP2}). However, in
both scenarios, the decay $H^{\pm}\to W^{\pm}A^{0}/H^{0}\to\ell^{\pm}E_{T}^{\mathrm{miss}}$
has a very small branching ratio, compared to $Br(H^{\pm}\to\ell^{\pm}N_{i})$,
which is about $8.27\%$ ($3.21\%$) for \textsf{BP1} (\textsf{BP2}).

\begin{table}[!t]
\centering \begin{adjustbox}{max width=\textwidth} %
\begin{tabular}{c|c|c}
\hline 
Benchmark Point & \multicolumn{1}{c|}{} & Parameters \tabularnewline
\hline 
\hline 
 & $m_{N_{i}}\left(\mathrm{GeV}\right)$ & $28.52,~30.97,~109$,\tabularnewline
\textsf{BP1} & $m_{H^{\pm}},m_{H^{0}}$ $\left(\mathrm{GeV}\right)$ & $164.2,~70.49$,\tabularnewline
 & $\lambda_{2},\lambda_{3},\lambda_{4},\lambda_{5}$ & $2.662,~-0.153,~-1.451,1.379\times10^{-10}$, \tabularnewline
 & $h_{ij}/10^{-2}$ & $\left(\begin{array}{ccc}
-14.218-i0.997 & -2.030-i0.999 & 62.265-i0.025\\
69.991+i0.438 & -0.544+i0.107 & -5.559+i0.719\\
-57.401+i0.980 & -0.188+i0.472 & -24.365-i0.419
\end{array}\right)$,\tabularnewline
\hline 
\hline 
 & $m_{N_{i}}\left(\mathrm{GeV}\right)$ & $17.93,~29.27,~54.49$,\tabularnewline
\textsf{BP2} & $m_{H^{\pm}},m_{H^{0}}$ $\left(\mathrm{GeV}\right)$ & $215.7,~124$,\tabularnewline
 & $\lambda_{2},\lambda_{3},\lambda_{4},\lambda_{5}$ & $1.069,~0.429,~-2.05,1.85\times10^{-10}$, \tabularnewline
 & $h_{ij}/10^{-2}$ & $\left(\begin{array}{ccc}
133.45+i5.739 & 0.239+i0.071 & -26.05+i0.57\\
0.79-i1.028 & -2.867+i2.957 & 78.53-i0.459\\
-68.17+i2.869 & 0.413+i0.082 & -57.06-i1.276
\end{array}\right)$.\tabularnewline
\hline 
\end{tabular}\end{adjustbox} \caption{The parameters values the benchmark points \textsf{BP1} and \textsf{BP2}.}
\label{benchmark.points} 
\end{table}

\begin{table}[!t]
\begin{centering}
\begin{tabular}{c|c|c|c|c|c}
\hline 
 & $\sigma(pp\to H^{\pm}H^{0}/A^{0})$ & $Br(W^{\pm}H^{0}/A^{0})$ & $\sum_{i}Br(e^{\pm}N_{i})$ & $\sum_{i}Br(\mu^{\pm}N_{i})$ & $\sum_{i}Br(\tau^{\pm}N_{i})$ \tabularnewline
\hline 
\hline 
\textsf{BP1} & $460.78$ fb & $37.22\%$ & $19.74\%$ & $24.07\%$ & $18.95\%$ \tabularnewline
\textsf{BP2} & $125.67$ fb & $14.46\%$ & $49.62\%$ & $16.25\%$ & $21.02\%$ \tabularnewline
\hline 
\end{tabular}
\par\end{centering}
\caption{$H^{\pm}S(S=A^{0},H^{0})$ production cross section and decays branching
ratios of the charged Higgs boson in the two benchmark scenarios.}
\label{decay.prod} 
\end{table}

\begin{figure}[!h]
\centering \includegraphics[width=0.5\linewidth]{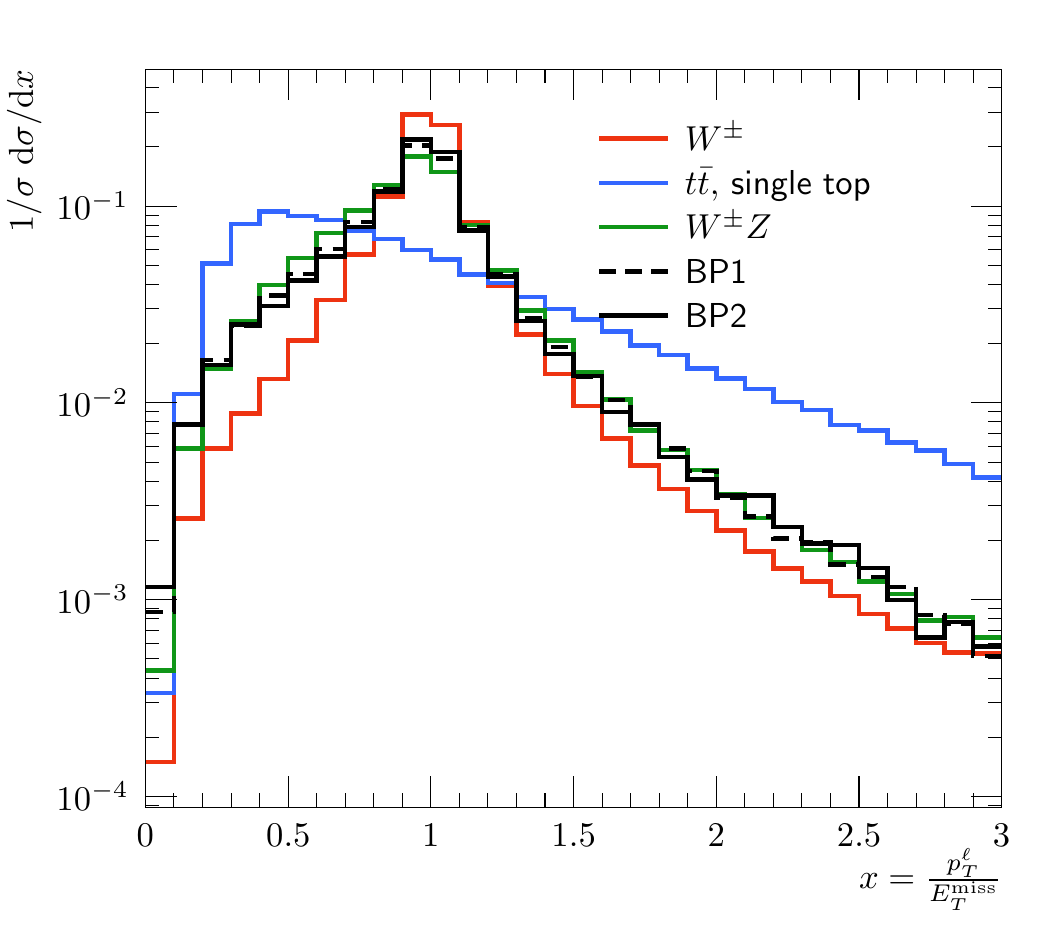} \caption{Transverse momentum imbalance between the charged lepton and the missing
energy in the two benchmark points and the SM background. The distribution
is normalized to unity.}
\label{pTMissET} 
\end{figure}

The background contributions to $\ell+E_{T}^{\mathrm{miss}}$ signal
can be classified into two categories: irreducible and reducible.
The $W$ production followed by its leptonic is the dominant irreducible
background with a cross section of order $\mathcal{O}(20)~nb$ at
Leading Order (LO). Diboson processes, $WZ,WW$ and $ZZ$, contribute
as well to the background sources in the $\ell+E_{T}^{\textrm{miss}}$
final state especially $W(\to\ell\nu)Z(\to\nu\bar{\nu})$ which is
irreducible. Important contributions might come from $t\bar{t}$ and
single top production where the top quark decays leptonically. Furthermore,
there are several background processes whose contribution cannot be
estimated at the parton level. In such background categories, charged
lepton and missing energy arise either from i) multi-jet production
where they are produced from hadron decays, and ii) in Drell-Yan process
($Z/\gamma^{*}\to\ell^{+}\ell^{-}$) where one lepton is not detected.
We do not consider these background in this study since their contribution
can be significantly reduced by imposing isolation cuts and by requiring
that the lepton and missing must have a back-to-back topology. These
requirements are translated into cuts on the ratio of $p_{T}^{\ell}/E_{T}^{\mathrm{miss}}$
and $\Delta\Phi(\ell,E_{T}^{\mathrm{miss}})$. A perfect balance in
the transverse plane between the charged lepton is reached when a
charged lepton is produced in association with an invisible particle
yielding a ratio $p_{T}^{\ell}/E_{T}^{\mathrm{miss}}\simeq1$ and
an azimuthal separation $\Delta\Phi\simeq\pi$. However, in hadronic
collisions, there are QCD radiation off partons in the proton beams
which result in a recoil of the lepton and missing energy system and
yield a shift in the values $p_{T}^{\ell}/E_{T}^{\mathrm{miss}}$
and $\Delta\Phi(\ell,E_{T}^{\mathrm{miss}})$ but with an unmodified
peak position. For reference, in Fig. \ref{pTMissET} we show the
imbalance between the lepton $p_{T}^{\ell}$ and transverse missing
energy in the two benchmark points and the SM background. We can see
that for the signal, $W^{\pm}$ and $W^{\pm}Z$, this distribution
is peaked around a value about $1$ while it is a smooth function
in the case of top quark processes.

\begin{figure}[!t]
\centering \includegraphics[width=0.48\linewidth]{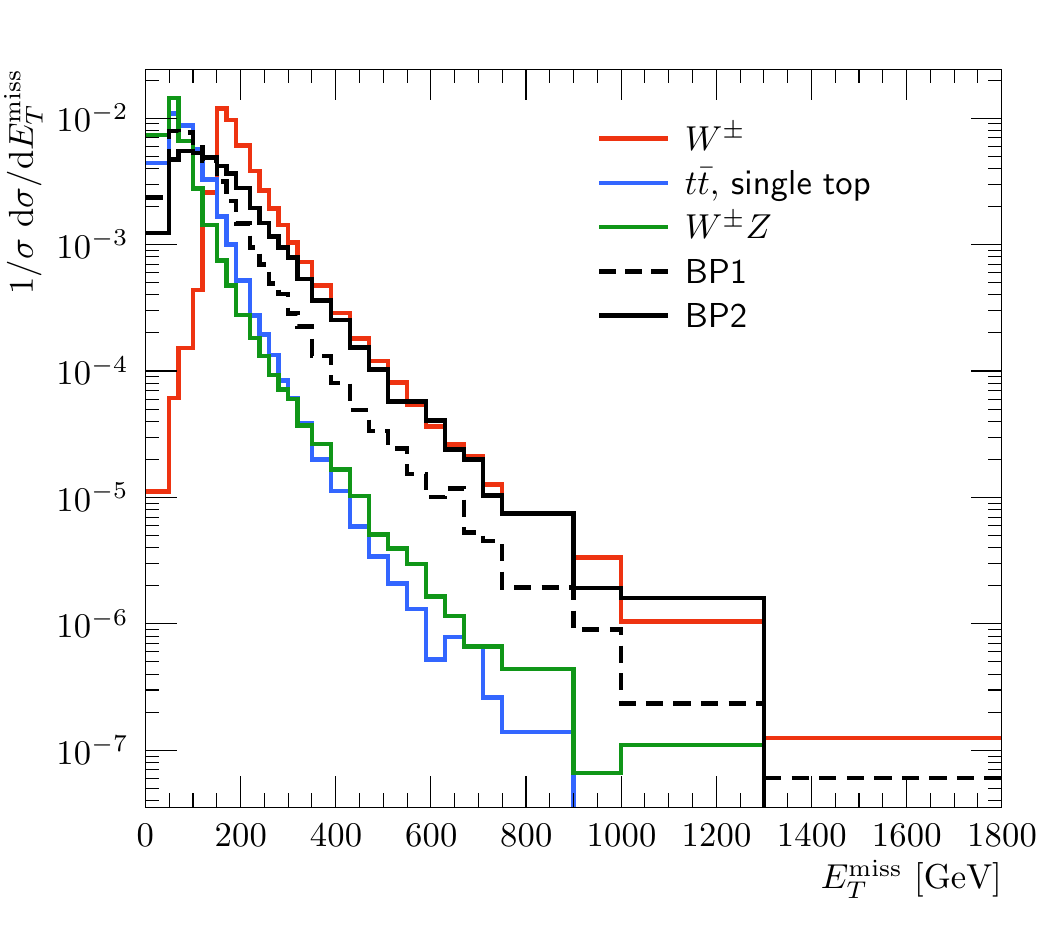} \hfill{}\includegraphics[width=0.48\linewidth]{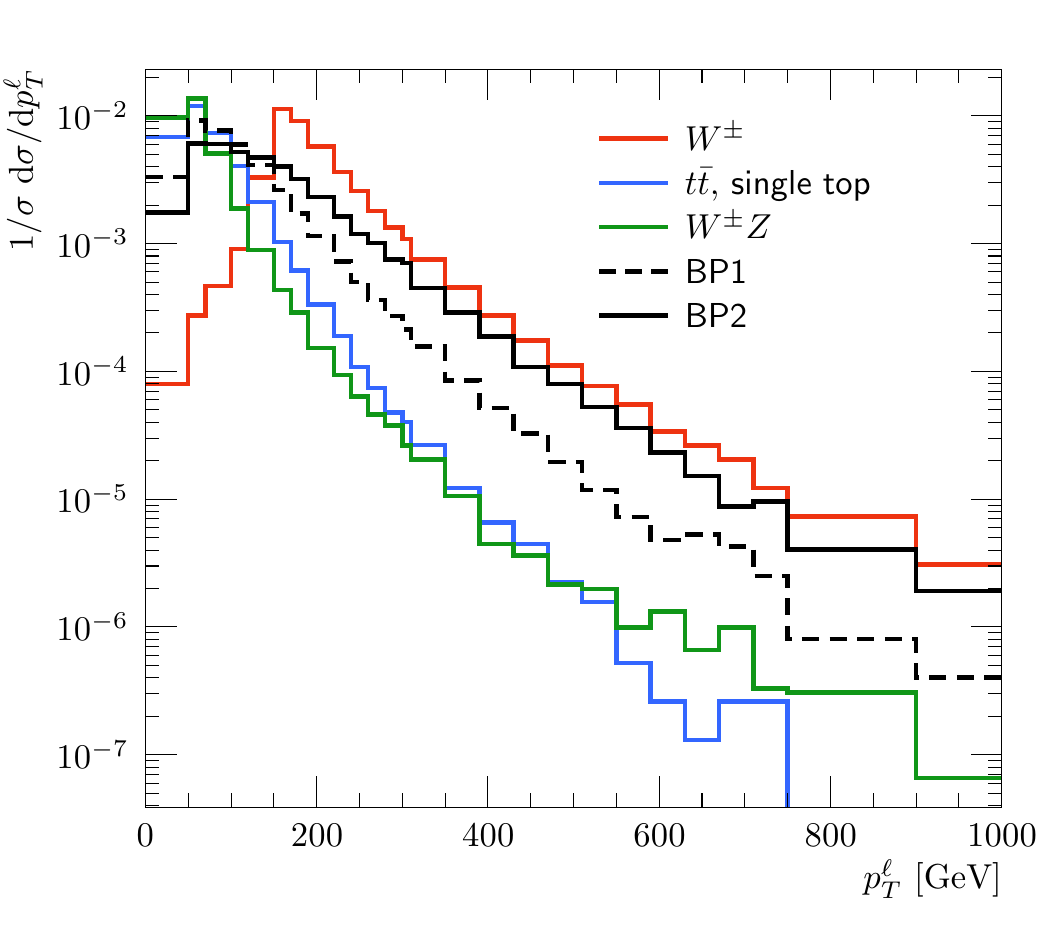}
\caption{Normalized distributions for missing transverse energy (left panel)
and lepton $p_{T}$ (right panel) for signal and background events.}
\label{missETpTL} 
\end{figure}

Signal events were generated at LO using \textsf{CalcHEP}~\cite{Belyaev:2012qa}
while $W^{\pm}$ and $W^{\pm}Z$ events were generated using \textsf{Madgraph5\_aMC@NLO}~\cite{Alwall:2011uj}.
\textsf{PYTHIA8}~\cite{Sjostrand:2007gs} was used for showering
and hadronization of the parton level events and for generation of
top quark events. Due to its large cross section, $W^{\pm}$ boson
events were generated with strong cuts, i.e with $p_{T}^{\ell}>100$
GeV and $E_{T}^{\textrm{miss}}>50$ GeV. Such cuts reduce the total
$W$ boson production cross section from $19$ nb to about $6$ pb.
While for the other background processes, no cuts were imposed on
the generated MC samples. At the analysis level, further cuts are
imposed on the charged lepton $p_{T}^{\ell}$ and $E_{T}^{\textrm{miss}}$
to further reduce the SM background contribution. We, first, select
events which contains exactly one charged lepton (either electron
or muon) with $p_{T}^{\ell}>30$ GeV and $|\eta|<2.4$, and a missing
transverse energy (MET) with $E_{T}^{\mathrm{miss}}>30$ GeV. Furthermore,
the imbalance between the lepton $p_{T}$ and MET was required to
be $0.4<p_{T}^{\ell}/E_{T}^{\mathrm{miss}}<1.4$. Such a cut reduce
the amount of events for the signal and the irreducible background
by about $20$-$25\%$. However, reducible backgrounds such as $t\bar{t}$
and single top quark production are reduced by a factor of $0.91$.
No cuts were imposed on the jet multiplicity, jet $p_{T}$ or $b$-tagging
since jet activity is involved in all the processes. We recommend
for a more complete analysis to be done at the experimental level.
We further refine our selection criteria and define two signal regions;
the first signal region is defined by $p_{T}^{\ell}>250$ GeV and
$E_{T}^{\mathrm{miss}}>150$ GeV while the second signal region is
defined as $p_{T}^{\ell}>250$ GeV and $E_{T}^{\mathrm{miss}}>300$
GeV. In Fig. \ref{missETpTL}, we show the corresponding distributions
for the signal as well as background events.

\begin{figure}[!t]
\centering \includegraphics[width=0.48\linewidth]{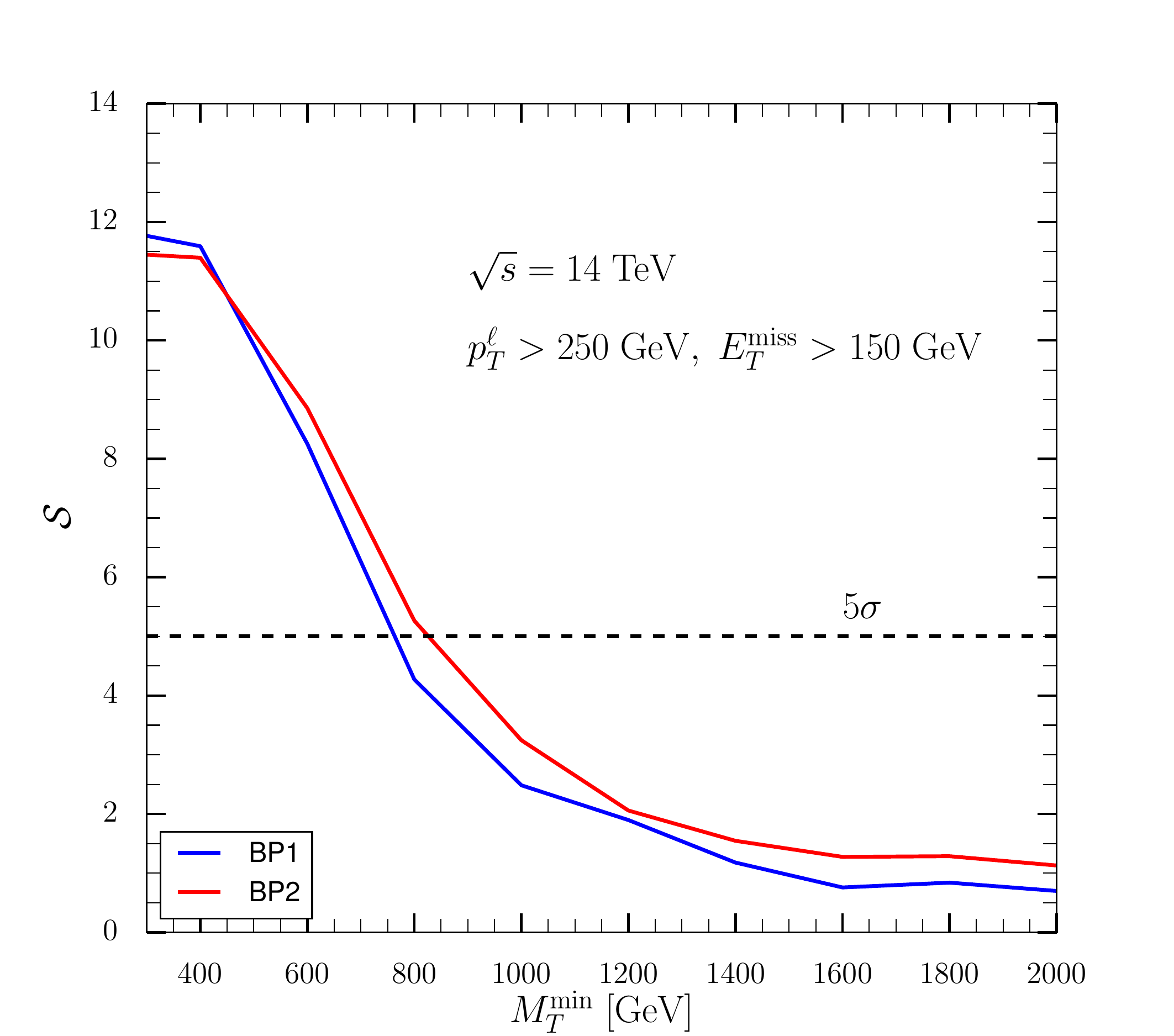}
\hfill{}\includegraphics[width=0.48\linewidth]{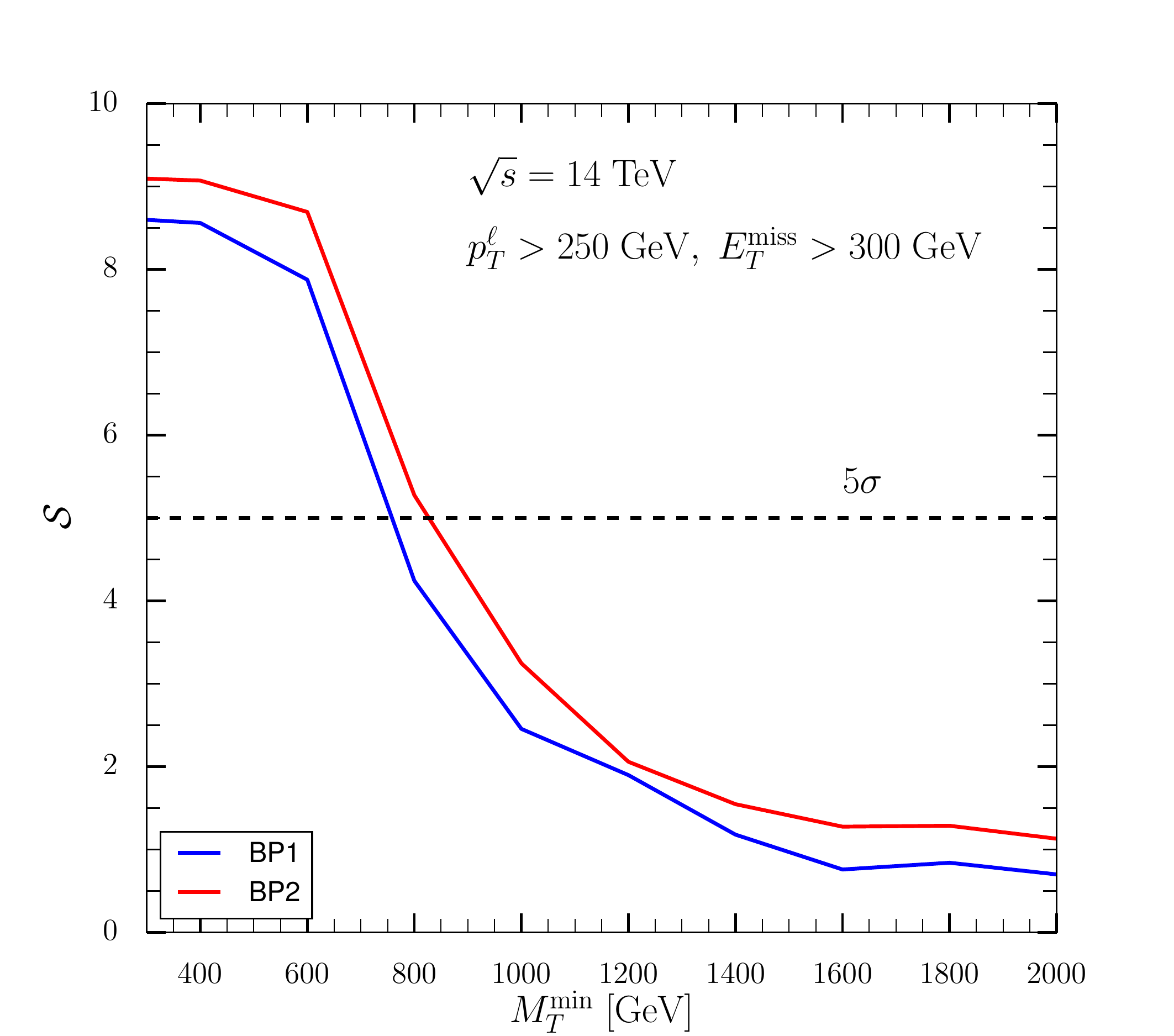}
\caption{Signal significance as function of $M_{T}^{\mathrm{min}}$ for the
first (left panel) and second (right panel) signal regions.}
\label{significance} 
\end{figure}

To quantify the potential discovery of the model in the two benchmark
points, we use the general formula for the signal significance which
is defined by 
\begin{equation}
\mathcal{S}=\sqrt{2\times\left[(N_{\textrm{S}}+N_{\textrm{BG}})\times\log\left(1+\frac{N_{\textrm{S}}}{N_{\textrm{BG}}}\right)-N_{\textrm{S}}\right]},\label{S}
\end{equation}
We compute $\mathcal{S}$ for different values of the cut on the transverse
mass of the lepton and missing transverse energy system defined by
\begin{eqnarray}
M_{T}=\sqrt{2p_{T}^{\ell}E_{T}^{\textrm{miss}}(1-\cos\Delta\Phi(\ell,E_{T}^{\textrm{miss}}))}\label{transverse-mass}
\end{eqnarray}
In Fig. \ref{significance}, we plot the significance as function
of the cut on $M_{T}$ for the two considered benchmark points and
in the two signal regions. For $M_{T}^{\mathrm{min}}<1$ TeV, the
significance reaches about $5\sigma$ level, and decreases quickly
for $M_{T}^{\mathrm{min}}>1$ TeV. This behavior at $M_{T}^{\mathrm{min}}>1$
TeV is due to the relatively low statistics at that region which is
a consequence of the light scalars chosen in our benchmark points.
On the other hand, benchmark points with heavier scalars hardly achieve
the $5\sigma$ level due to the smallness of the corresponding cross
section. We notice that in the observability region, systematic uncertainties
are quite moderate. They are dominated by uncertainties due to electron
and muon energy resolution. Statistical uncertainties are also not
important in this region of interest. We found that even for lower
luminosities of about $100$ fb$^{-1}$, the significance can reach
the $5\sigma$ level for $M_{T}^{\mathrm{min}}<600$ GeV in \textsf{BP1}.
We leave a more detailed study of all the systematic uncertainties
in mono-lepton signatures and interplay with the other channels for
a future work ~\cite{Ahriche:2018}. 

\section{Conclusion}

\label{sec:conclusion}

In this paper, we considered the inert Higgs Doublet Model extended
by three right handed neutrinos where both neutrino mass and dark
matter are addressed. We considered the DM to be the lightest right
handed neutrino and showed that its relic density can be in agreement
with the observation provided that Yukawa couplings in the neutrino
sector are not highly suppressed. We carried out a detailed numerical
analysis to determine the different regions of the parameter space
that are consistent with theoretical and experimental constraints.
Fitting the observed neutrino mass squared differences and mixing
angles requires that the CP-odd and the CP-even components of the
inert doublet to have quasi-degenerate masses. We also discussed a
number of experimental signatures of this model at high energy colliders.
In particular, mono-lepton, mono-jet and mono-photon signals are particularly
interesting and can be used to search for the right handed neutrinos
both at the LHC and future lepton colliders, such as the ILC. We have
performed at detailed analysis of the mono-lepton signature at the
LHC and showed that with a $300~fb^{-1}$ luminosity it is possible
to probe the right handed neutrino signal.

\appendix
%dummy comment inserted by tex2lyx to ensure that this paragraph is not empty

\section{The cross section of $N_{i}-N_{k}$ Co-annihilation}

\begin{figure}[!h]
\begin{centering}
\includegraphics[width=12cm,height=3.5cm]{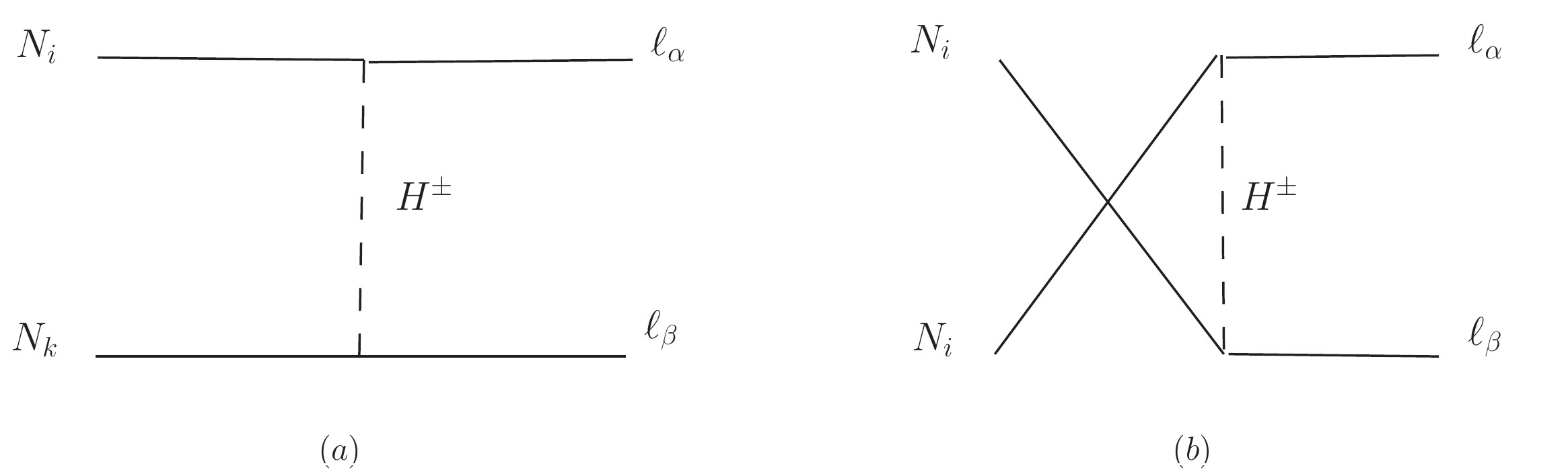} 
\par\end{centering}
\caption{\textit{Diagrams contributing to the annihilation $N_{i}N_{k}\rightarrow\ell_{\alpha}^{-}\ell_{\beta}^{+}$.
Clearly, the second diagram is possible only if the two RHN's are
identical $(i=k)$. Similar diagrams exit for the channels $N_{i}N_{k}\rightarrow\nu_{\alpha}\bar{\nu}_{\beta}$,
where the mediators are $H^{0}/A^{0}$. }}
\label{NNll} 
\end{figure}

In this appendix, we derive the analytic expression annihilation cross
of two right handed neutrinos, $N_{i}$ and $N_{k}$, into charged
leptons or light neutrinos. For the charged leptons channel, there
are two diagrams that contribute to this process, as shown in Fig.~\ref{NNll}.

After averaging and summing over the initial and final spin states,
the corresponding squared amplitude is given by

\begin{align}
\left\vert \mathcal{M}\right\vert ^{2} & =\frac{\left\vert g_{i\alpha}g_{k\beta}^{\ast}\right\vert ^{2}}{4}\left[\tfrac{\left(M_{i}^{2}+m_{\ell_{\alpha}}^{2}-t\right)\left(M_{k}^{2}+m_{\ell_{\beta}}^{2}-t\right)}{\left(t-m_{H^{+}}^{2}\right)^{2}}+\delta_{ik}\tfrac{\left(M_{i}^{2}+m_{\ell_{\alpha}}^{2}-u\right)\left(M_{i}^{2}+m_{\ell_{\beta}}^{2}-u\right)}{\left(u-m_{H^{+}}^{2}\right)^{2}}-\delta_{ik}\tfrac{2M_{i}^{2}\left(s-m_{\ell_{\alpha}}^{2}-m_{\ell_{\beta}}^{2}\right)}{\left(t-m_{H^{+}}^{2}\right)\left(u-m_{H^{+}}^{2}\right)}\right],
\end{align}
with $s,t,u$ and $s$ are the Lorentz invariant Mandelstam variables.

After integrating over the phase space, the cross section times the
relative velocity of $N_{i}$ and $N_{k}$ reads 
\begin{align}
\upsilon_{r}\sigma & =\frac{\left\vert g_{i\alpha}g_{k\beta}^{\ast}\right\vert ^{2}}{32\pi}\frac{\sqrt{s^{2}-2s\left(m_{\ell_{\alpha}}^{2}+m_{\ell_{\beta}}^{2}\right)+\left(m_{\ell_{\alpha}}^{2}-m_{\ell_{\beta}}^{2}\right)^{2}}}{s^{2}-\left(M_{i}^{2}-M_{k}^{2}\right)^{2}}\left[\tfrac{A_{1}^{2}-B^{2}+Q_{1}Q_{2}}{A_{1}^{2}-B^{2}}-\frac{Q_{1}+Q_{2}}{2B}\log\left|\frac{A_{1}+B}{A_{1}-B}\right|\right|+\delta_{ik}\times\nonumber \\
 & \left.\left\{ \tfrac{A_{2}^{2}-B^{2}+Q_{1}Q_{2}}{A_{2}^{2}-B^{2}}-\frac{Q_{1}+Q_{2}}{2B}log\left|\frac{A_{2}+B}{A_{2}-B}\right|-\frac{M_{i}^{2}\left(s-m_{\ell_{\alpha}}^{2}-m_{\ell_{\beta}}^{2}\right)}{B\left(Q_{1}+Q_{2}-s\right)}\left(\log\left|\frac{A_{1}+B}{A_{1}-B}\right|+\log\left|\frac{A_{2}+B}{A_{2}-B}\right|\right)\right\} \right],\label{eq:Sig}
\end{align}
where 
\begin{align*}
Q_{1} & =M_{i}^{2}+m_{\ell_{\alpha}}^{2}-m_{H^{+}}^{2},~Q_{2}=M_{k}^{2}+m_{\ell_{\beta}}^{2}-m_{H^{+}}^{2},\\
A_{1} & =\frac{M_{i}^{2}+M_{k}^{2}+m_{\ell_{\alpha}}^{2}+m_{\ell_{\beta}}^{2}-2m_{H^{+}}^{2}-s}{2}-\frac{\left(M_{i}^{2}-M_{k}^{2}\right)\left(m_{\ell_{\alpha}}^{2}-m_{\ell_{\beta}}^{2}\right)}{2s},\\
A_{2} & =M_{i}^{2}-m_{H^{+}}^{2}+\frac{m_{\ell_{\alpha}}^{2}+m_{\ell_{\beta}}^{2}-s}{2},\,B=2\frac{\sqrt{s^{2}-2s\left(m_{\ell_{\alpha}}^{2}+m_{\ell_{\beta}}^{2}\right)+\left(m_{\ell_{\alpha}}^{2}-m_{\ell_{\beta}}^{2}\right)^{2}}}{2\sqrt{s}}p.
\end{align*}
Here $p=\sqrt{s^{2}-2s\left(M_{i}^{2}+M_{k}^{2}\right)+\left(M_{i}^{2}-M_{k}^{2}\right)^{2}}/2\sqrt{s}$
is the magnitude of the momentum of the incoming RHN in the center
of mass frame. The cross section of $\sigma(N_{i}N_{k}\rightarrow\nu_{\alpha}\nu_{\beta}$
can be obtained by simply making the replacement $~m_{H^{+}}^{2}\rightarrow\bar{m}^{2}=\frac{1}{2}(m_{H^{0}}^{2}+m_{A^{0}}^{2})$
and $m_{\ell_{\alpha}}^{2}=m_{\ell_{\beta}}^{2}=0,$ in the above
expression (\ref{eq:Sig}).

In the non-relativistic limit, we can expand the logarithms in (\ref{eq:Sig})
in power of $p$, equivalently $B$, to obtain 
\begin{align*}
\upsilon_{r}\sigma & \approx\frac{\left\vert g_{i\alpha}g_{k\beta}^{\ast}\right\vert ^{2}}{32\pi}\frac{\sqrt{s^{2}-2s\left(m_{\ell_{\alpha}}^{2}+m_{\ell_{\beta}}^{2}\right)+\left(m_{\ell_{\alpha}}^{2}-m_{\ell_{\beta}}^{2}\right)^{2}}}{s^{2}-\left(M_{i}^{2}-M_{k}^{2}\right)^{2}}\left[\frac{A_{1}-Q-Q_{2}}{A_{1}}+\tfrac{Q_{1}Q_{2}}{A_{1}^{2}}\left(1+\frac{B^{2}}{A_{1}^{2}}\right)\right.\\
 & \left.+\delta_{ik}\left\{ \frac{A_{2}-Q_{1}-Q_{2}}{A_{2}}+\tfrac{Q_{1}Q_{2}}{A_{2}^{2}}\left(1+\frac{B^{2}}{A_{2}^{2}}\right)-\frac{2M_{i}^{2}\left(s-m_{\ell_{\alpha}}^{2}-m_{\ell_{\beta}}^{2}\right)}{A_{2}^{2}}\right\} \right].
\end{align*}
For $i=k=1$, the expression of the annihilation cross section agrees
with the one given in~\cite{Cheung:2004xm}.

\section{The effective $hN_{1}\bar{N}_{1}$ coupling}

\label{appendix2}

Here, we present the different contributions to the effective $hN_{1}\bar{N}_{1}$
coupling in terms of the Passarino-Veltman three-point functions (denoted
by $C_{i}$ in what follows). To compute the effective $\tilde{y}_{hN_{1}\bar{N}_{1}}$
coupling, we consider the process $h(p)\to N_{1}(k_{1})+\bar{N}_{1}(k_{2})$,
at one-loop level. The corresponding Feynman diagrams are depicted
in Fig.~\ref{fig:hNN}. The evaluation of the Feynman amplitudes
was performed analytically and compared to the result of \textsf{FeynArts}
and \textsf{FormCalc}~\cite{FA2} while numerical evaluations of
the coupling were performed with the help of the \textsf{LoopTools}
package~\cite{FF} and compared to the approximate result shown in
(\ref{yhNN}). The amplitude of the first contribution (diagram (a)
in Fig.~\ref{fig:hNN}) can be written as 
\begin{eqnarray}
i\Gamma_{a}\left(p,k_{1},k_{2}\right) & = & \int\frac{d^{4}Q}{\left(2\pi\right)^{4}}\left(-ih_{\alpha1}^{*}P_{L}\right)\frac{i}{\slashed{Q}-m_{\ell_{\alpha}}}\left(i\frac{m_{\ell_{\alpha}}}{\upsilon}\right)\frac{i}{\slashed{Q}+\slashed{p}-m_{\ell_{\alpha}}}\left(ih_{\alpha1}P_{R}\right)\frac{i}{\left(k_{1}-Q\right)^{2}-m_{H^{+}}^{2}}\\
 & = & h_{\alpha1}^{*}h_{\alpha1}\frac{m_{\ell_{\alpha}}}{\upsilon}\int\frac{d^{4}Q}{\left(2\pi\right)^{4}}\frac{P_{L}\left(\slashed{Q}+m_{\ell_{\alpha}}\right)\left(\slashed{Q}+\slashed{p}+m_{\ell_{\alpha}}\right)P_{R}}{\left[Q^{2}-m_{\ell_{\alpha}}^{2}\right]\left[\left(Q-p\right)^{2}-m_{\ell_{\alpha}}^{2}\right]\left[\left(k_{1}-Q\right)^{2}-m_{H^{+}}^{2}\right]}\nonumber \\
 & = & \frac{h_{\alpha1}^{*}h_{\alpha1}m_{\ell_{\alpha}}^{2}}{\upsilon}P_{L}\int\frac{d^{4}Q}{\left(2\pi\right)^{4}}\frac{2\slashed{Q}+\slashed{p}}{\left[Q^{2}-m_{\ell_{\alpha}}^{2}\right]\left[\left(Q-p\right)^{2}-m_{\ell_{\alpha}}^{2}\right]\left[\left(k_{1}-Q\right)^{2}-m_{H^{+}}^{2}\right]}.
\end{eqnarray}
Using the Passarino-Veltman reduction~\cite{Passarino:1978jh}, we
find for the first contribution
\begin{align}
\Gamma_{a}\left(p,k_{1},k_{2}\right) & =\frac{1}{16\pi^{2}\upsilon}\sum_{\alpha}|h_{\alpha1}|^{2}m_{\ell_{\alpha}}^{2}\left\{ C_{0}(M_{1}^{2},p^{2},M_{1}^{2},m_{H^{\pm}}^{2},m_{\ell_{\alpha}}^{2},m_{\ell_{\alpha}}^{2}) \right.\nonumber \\
 & +\left.C_{1}(M_{1}^{2},p^{2},M_{1}^{2},m_{H^{\pm}}^{2},m_{\ell_{\alpha}}^{2},m_{\ell_{\alpha}}^{2})+C_{2}(M_{1}^{2},p^{2},M_{1}^{2},m_{H^{\pm}}^{2},m_{\ell_{\alpha}}^{2},m_{\ell_{\alpha}}^{2})\right\} .
\end{align}

The amplitude for the second contribution (diagram (b) in Fig.~\ref{fig:hNN})
is given by 
\begin{align}
i\Gamma_{b}\left(p,k_{1},k_{2}\right) & =\int\frac{d^{4}Q}{\left(2\pi\right)^{4}}\left(-ih_{\alpha1}^{*}P_{L}\right)\frac{i}{\slashed{k}_{1}-\slashed{Q}-m_{\ell_{\alpha}}}\left(ih_{\alpha1}P_{R}\right)\frac{i}{\left(p+Q\right)^{2}-m_{H^{+}}^{2}}\left(i\lambda_{3}\upsilon\right)\frac{i}{Q^{2}-m_{H^{+}}^{2}}\nonumber \\
 & +(P_{L}\longleftrightarrow P_{R}).
\end{align}
After some algebra, we get
\begin{align}
\Gamma_{b}\left(p,k_{1},k_{2}\right) & =-\frac{1}{16\pi^{2}}\lambda_{3}\upsilon M_{1}\sum_{\alpha}|h_{\alpha1}|^{2}\left\{ C_{0}(p^{2},M_{1}^{2},M_{1}^{2},m_{H^{\pm}}^{2},m_{H^{\pm}}^{2},m_{\ell_{\alpha}}^{2})\right.\nonumber \\
 & +\left.C_{2}(p^{2},M_{1}^{2},M_{1}^{2},m_{H^{\pm}}^{2},m_{H^{\pm}}^{2},m_{\ell_{\alpha}}^{2})\right\} .
\end{align}

The contribution of diagram (c) can be evaluated similarly to give
\begin{align}
\Gamma_{c}\left(p,k_{1},k_{2}\right) & =-\frac{1}{16\pi^{2}}(\lambda_{3}+\lambda_{4}/2)M_{1}\upsilon\sum_{S=H^{0},A^{0}}\sum_{\alpha}|h_{\alpha1}|^{2}\left\{ C_{1}(p^{2},M_{1}^{2},M_{1}^{2},0,m_{S}^{2},m_{S}^{2})\right.\nonumber \\
 & +\left.C_{2}(p^{2},M_{1}^{2},M_{1}^{2},0,m_{S}^{2},m_{S}^{2})\right\}.
\end{align}
Numerically, we
found that the contribution of diagram (a) is much smaller than the
contribution of diagrams (b) and (c). Furthermore, in our numerical
analysis, we use the expression quoted in (\ref{yhNN}) since it agrees
very well with the full expression in terms of the Passarino-Veltman
functions.

\acknowledgments

We want to thank M. Chekkal for his help in the results production
in Fig.~\ref{fig:charged}. The work of AJ was supported by Shanghai
Pujiang Program and by the Moroccan Ministry of Higher Education and
Scientific Research MESRSFC and CNRST: `` Projet dans
les domaines prioritaires de la recherche scientifique et du developpement
technologique`` : PPR/2015/6.

\end{document}